\pdfoutput=1
\documentclass[prd,onecolumn,notitlepage,eqsecnum,nofootinbib,floatfix]{revtex4-1}
\usepackage{amssymb}
\usepackage{amsmath}
\usepackage{bm}
\usepackage{color} 
\usepackage{graphicx}
\DeclareFontFamily{OT1}{rsfs}{} 
\DeclareFontShape{OT1}{rsfs}{m}{n}{<-7> rsfs5 
    <7-10> rsfs7 <10-> rsfs10}{}   
\DeclareMathAlphabet{\scr}{OT1}{rsfs}{m}{n} 

\newcommand{\E}{{\cal E}} 
\newcommand{\B}{{\cal B}} 
\newcommand{\F}{{\cal F}} 
\newcommand{\K}{{\cal K}} 
\newcommand{\N}{{\scr N}} 
\newcommand{\chid}{\chi^{\scriptstyle \sf d}} 
\newcommand{\Eq}{{\cal E}^{\scriptstyle \sf q}} 
\newcommand{\Edotq}{\dot{\cal E}^{\scriptstyle \sf q}} 
\newcommand{\Bq}{{\cal B}^{\scriptstyle \sf q}} 
\newcommand{\Fd}{{\cal F}^{\scriptstyle \sf d}} 
\newcommand{\Fo}{{\cal F}^{\scriptstyle \sf o}} 
\newcommand{\Kd}{{\cal K}^{\scriptstyle \sf d}} 
\newcommand{\Ko}{{\cal K}^{\scriptstyle \sf o}} 
\newcommand{\m}{{\sf m}} 
\newcommand{\eq}{e^{\scriptstyle \sf q}} 
\newcommand{\bq}{b^{\scriptstyle \sf q}} 
\newcommand{\ehatq}{\hat{e}^{\scriptstyle \sf q}} 
\newcommand{\ebarq}{\bar{e}^{\scriptstyle \sf q}} 
\newcommand{\ebarhatq}{\hat{\bar{e}}^{\scriptstyle \sf q}} 
\newcommand{\bhatq}{\hat{b}^{\scriptstyle \sf q}} 
\newcommand{\kd}{k^{\scriptstyle \sf d}} 
\newcommand{\kbard}{\bar{k}^{\scriptstyle \sf d}} 
\newcommand{\ko}{k^{\scriptstyle \sf o}} 
\newcommand{\kbaro}{\bar{k}^{\scriptstyle \sf o}} 
\newcommand{\fd}{f^{\scriptstyle \sf d}} 
\newcommand{\fo}{f^{\scriptstyle \sf o}} 
\newcommand{\cc}[1]{c^{\scriptstyle \sf #1}} 
\newcommand{\gam}[1]{\gamma^{\scriptstyle \sf #1}} 
\newcommand{\pd}{p^{\scriptstyle \sf d}} 
\newcommand{\pq}{p^{\scriptstyle \sf q}} 
\newcommand{\po}{p^{\scriptstyle \sf o}} 
\newcommand{\qd}{q^{\scriptstyle \sf d}} 
\newcommand{\qq}{q^{\scriptstyle \sf q}} 
\newcommand{\qo}{q^{\scriptstyle \sf o}} 
\newcommand{\PN}{\mbox{\sc pn}} 
\allowdisplaybreaks
\begin{document}
\title{Tidal deformation of a slowly rotating black hole} 
\author{Eric Poisson} 
\affiliation{Department of Physics, University of Guelph, Guelph,
  Ontario, N1G 2W1, Canada} 
\affiliation{$\mathcal{G}\mathbb{R}\varepsilon{\mathbb{C}}\mathcal{O}$,
  Institut d'Astrophysique de Paris, 98\textsuperscript{bis}
  boulevard Arago, 75014 Paris, France}
\date{January 5, 2015} 
\begin{abstract} 
In the first part of this article I determine the geometry of a slowly
rotating black hole deformed by generic tidal forces created by a
remote distribution of matter. The metric of the deformed black hole
is obtained by integrating the Einstein field equations in a vacuum
region of spacetime bounded by $r < r_{\rm max}$, with $r_{\rm max}$ a
maximum radius taken to be much smaller than the distance to the
remote matter. The tidal forces are assumed to be weak and to vary
slowly in time, and the metric is expressed in terms of generic tidal
quadrupole moments $\E_{ab}$ and $\B_{ab}$ that characterize the tidal
environment. The metric incorporates couplings between the black
hole's spin vector and the tidal moments, and captures all effects
associated with the dragging of inertial frames. In the second 
part of the article I determine the tidal moments by immersing the
black hole in a larger post-Newtonian system that includes an external  
distribution of matter; while the black hole's internal gravity is
allowed to be strong, the mutual gravity between the black hole and
the external matter is assumed to be weak. The post-Newtonian metric
that describes the entire system is valid when $r > r_{\rm min}$,
where $r_{\rm min}$ is a minimum distance that must be much larger
than the black hole's radius. The black-hole and post-Newtonian
metrics provide alternative descriptions of the same gravitational
field in an overlap $r_{\rm min} < r < r_{\rm max}$, and matching the
metrics determine the tidal moments, which are expressed as
post-Newtonian expansions carried out through one-and-a-half
post-Newtonian order. Explicit expressions are obtained in the
specific case in which the black hole is a member of a post-Newtonian
two-body system.     
\end{abstract} 
\maketitle

\section{Introduction and summary} 
\label{sec:intro} 

\subsection*{This work and its context} 

The theory of tidal deformation and dynamics of compact bodies in
general relativity has recently been the subject of vigorous
development. While work on this topic goes back several decades, the
origin of the recent burst of activity can be traced to Flanagan and
Hinderer \cite{flanagan-hinderer:08, hinderer:08}, who pointed out
that tidal effects can have measurable consequences on the
gravitational waves emitted by a binary neutron star in the late 
stages of its orbital evolution. The gravitational waves therefore
carry information regarding the internal structure of each body, 
from which one can extract useful constraints on the equation of state
of dense nuclear matter. Their study was followed up with more
detailed analyses \cite{hinderer-etal:10, baiotti-etal:10,
  baiotti-etal:11, vines-flanagan-hinderer:11, pannarale-etal:11,
  lackey-etal:12, damour-nagar-villain:12, read-etal:13,
  vines-flanagan:13, lackey-etal:14, favata:14, yagi-yunes:14}, with
the conclusion that tidal effects might indeed be accessible to
measurement by the current generation of gravitational-wave detectors.  

Another astrophysical context for relativistic tidal dynamics comes
from extreme-mass-ratio inspirals, which implicate a solar-mass
compact body in a tight orbit around a supermassive black hole. Such
systems have been identified as promising sources of gravitational
waves for an eventual space-borne interferometric detector, and it was
demonstrated \cite{hughes:01, price-whelan:01, martel:04,
  yunes-etal:10, yunes-etal:11, chatziioannou-poisson-yunes:13} that
the tide raised on the large black hole by the small body can lead to
a significant transfer of angular momentum from the black hole to the
orbital motion.    

These observations have motivated the formulation of a relativistic
theory of tidal deformation and dynamics that can be applied to
neutron stars and black holes. The description of tidal deformations,
featuring the relativistic generalization of the Newtonian Love
numbers \cite{damour-nagar:09, binnington-poisson:09,
  landry-poisson:14}, is now mature, and the relativistic Love numbers
have been computed for realistic neutron-star models constructed from
tabulated equations of state \cite{hinderer-etal:10, postnikov-etal:10,
  pannarale-etal:11, damour-nagar-villain:12}; the gravitational Love 
numbers of a black hole were shown to be precisely zero 
\cite{binnington-poisson:09}. The Love numbers of neutron stars have
been implicated in a remarkable set of nearly universal relations ---
the $I$-Love-$Q$ relations \cite{yagi-yunes:13a, yagi-yunes:13b,
  doneva-etal:14, maselli-etal:13, yagi:14, haskell-etal:14,
  chakrabarti-etal:14}  --- involving the moment of inertia $I$, the
Love number $k_2$, and the rotational quadrupole moment $Q$ of a
neutron star.     

In another development, the tidal deformation of compact bodies has
been incorporated in an effective description in terms of point
particles; the description involves a point-particle action that
includes invariants constructed from the tidal multipole moments to be 
introduced below. In Ref.~\cite{bini-damour-faye:12}, Bini, Damour,
and Faye expressed the tidal invariants as a post-Newtonian expansion
carried out through second post-Newtonian order. In a recent update 
\cite{bini-damour:14}, Bini and Damour pushed the expansion to
seven-and-a-half post-Newtonian order (restricted to small mass
ratios), and showed that their results compare well with numerical
calculations of tidal invariants obtained from the gravitational
self-force \cite{dolan-etal:14}; the work by Bini and Damour extends 
a previous study by Chakrabarti, Delsate, and Steinhoff 
\cite{chakrabarti-delsate-steinhoff:13a} (see also
Ref.~\cite{chakrabarti-delsate-steinhoff:13b}). 

This work is an additional contribution to the development of a
relativistic theory of tidal deformations. I specifically consider
the tidal deformation of a black hole, and in the first part of the
paper I determine the geometry of the deformed black hole in terms of 
generic tidal moments $\E_{ab}$ and $\B_{ab}$; in the second
part I determine the tidal moments by immersing the black hole in a
post-Newtonian environment. The novelty of this work lies in the
fact that I allow the black hole to be slowly rotating, so as to
capture all effects associated with the dragging of inertial frames,
which are included in couplings between the black-hole spin 
vector and the tidal moments. The assumption of slow rotation allows
me to neglect the rotational deformation of the black hole.     

\subsection*{Geometry of a slowly rotating, tidally deformed black
  hole}  

\begin{figure} 
\includegraphics[width=0.5\linewidth]{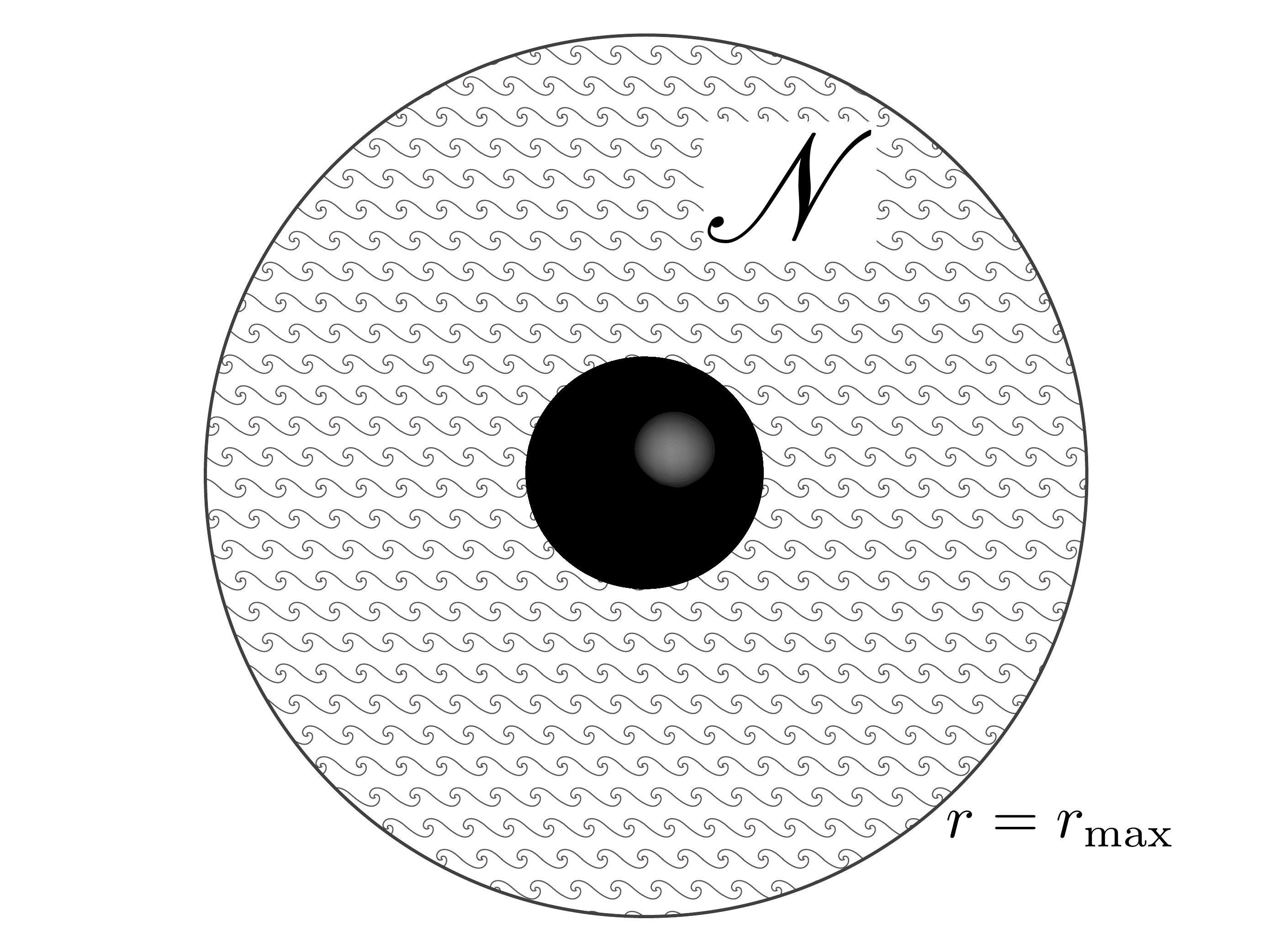}
\caption{The domain $\N$ around the black hole, bounded by 
  $r < r_{\rm max}$, is shown in a wavy pattern. The external matter,
  situated at a distance $b \gg r_{\rm max}$ far outside the domain,
  is not shown.}     
\label{fig:f1} 
\end{figure}

My goal in the first part of this article is to determine the
geometry of a slowly rotating black hole deformed by generic tidal
forces exerted by remote matter. I assume that the tidal forces are
weak, that they vary slowly in time, and use perturbative methods to
integrate the Einstein field equations for this situation.  I focus
my attention on a domain $\N$ bounded by $r < r_{\rm max}$ (see 
Fig.~\ref{fig:f1}), with $r$ denoting the distance to the black hole,
and $r_{\rm max} \ll b$ a maximum distance that is taken to be much
smaller than $b$, the distance to the external matter; the domain is
assumed to be empty of matter. I construct the metric of the deformed
black hole in $\N$, and express it as an expansion in powers of
$r/b$. Because $\N$ does not include the external matter, the metric
is expressed in terms of tidal moments $\E_{ab}$ and $\B_{ab}$ that
cannot be determined by the vacuum field equations restricted to $\N$.    

Such constructions originate in the seminal work of Manasse
\cite{manasse:63}, which was revived by Alvi \cite{alvi:00, alvi:03},
Detweiler \cite{detweiler:01, detweiler:05}, and this author
\cite{poisson:05, poisson-vlasov:10}. The geometry of a tidally
deformed, rapidly rotating black hole was described by Yunes and
Gonzalez \cite{yunes-gonzalez:06}, and more recently by O'Sullivan and
Hughes \cite{o'sullivan-hughes:14}. While Yunes and Gonzalez obtain an
analytical expression for the metric in $\N$ by integrating the
Teukolsky equation (for weak and slow tides) and exploiting 
metric-reconstruction techniques, O'Sullivan and Hughes focus on the
intrinsic geometry of the event horizon, which they determine
numerically for the rapidly changing tidal forces produced by an
orbiting body. Because I take the black hole to be rotating slowly,
the calculations presented here are a simplification of the work
carried out by Yunes and Gonzalez. What is gained from this
simplification is an explicit expression for the metric that can be
displayed in a few lines, a metric cast in a coordinate system that
possesses a clear geometrical meaning, and much more insight into the
coupling between rotational and tidal effects.  

It is helpful to introduce the various scales that enter the
description of the black hole and its tidal environment; I work in
geometrized units, in which $G=c=1$. The black-hole mass is $m_1$, and
its spin vector $S^a_1$ is related to a dimensionless quantity
$\chi^a_1$ by $S^a_1 = \chi^a_1 m_1^2$. The Kerr solution describes a
black hole when $\chi_1 := |\chi^a_1| \leq 1$, but here I demand that 
$\chi_1 \ll 1$, so that the black hole is rotating slowly. The
external matter is characterized by a mass scale $m_2$ and its
distance to the black hole is comparable to the length scale $b$. The
time scale associated with changes in the tidal environment is 
$\tau \sim \sqrt{b^3/m}$, in which $m := m_1 + m_2$ is a scale for the
total mass of the system. A velocity scale is then 
$u \sim b/\tau = \sqrt{m/b}$. The tidal environment is characterized
by an electric-type tidal quadrupole moment $\E_{ab}$ that scales as
$m_2/b^3$, and a magnetic-type tidal quadrupole moment $\B_{ab}$ that
scales as $m_2 u/b^3$. The leading tidal terms in the metric are
proportional to $r^2 \E_{ab} \sim (m_2/b) (r/b)^2$ and 
$r^2 \B_{ab} \sim (m_2/b) u (r/b)^2$; these are small in $\N$ by
virtue of the fact that $r \ll b$. The next-to-leading tidal terms
involve the time derivative of the tidal quadrupole moments as well as
tidal octupole moments, and those are suppressed by additional factors  
of $u(r/b)$ and $(r/b)$, respectively. Here I assume that the
coupling terms between $\chi^a_1$ and the tidal quadrupole moments, 
which are proportional to $\chi^a_1 \E_{bc}$ and $\chi^a_1 \B_{bc}$
and therefore suppressed relative to the leading terms by factors of
order $\chi_1$, nevertheless dominate over the next-to-leading
terms. This requires $\chi_1 \gg (r/b)$, and taking $r$ to be of the
same order of magnitude as $m$ in $\N$, we find that the spin
parameter must be constrained by $u^2 \ll \chi_1 \ll 1$. Because 
$u^2 \ll 1$, the mutual gravity between the black hole and the
external matter is required to be weak.   

The tidal potentials that result from the couplings between the spin
vector and the tidal moments are introduced in 
Sec.~\ref{sec:potentials}. The background spacetime of a slowly
rotating black hole in isolation is described in
Sec.~\ref{sec:background}, and its metric is cast in coordinates
$(v,r,\theta,\phi)$ that are tied to the behavior of incoming null
geodesics that are tangent to converging null cones. In
Sec.~\ref{sec:perturbed} I introduce the tidal deformation,
characterized by tidal moments $\E_{ab}(v)$ and $\B_{ab}(v)$, and I 
integrate the Einstein field equations to find the metric of the
deformed black hole in the domain $\N$. I adopt a gauge 
--- the light-cone gauge \cite{preston-poisson:06b} --- that preserves 
the geometrical meaning of the $(v,r,\theta,\phi)$ coordinates, so
that $v$ continues to be constant on converging null cones, $\theta$
and $\phi$ continue to be constant on each generator, and $-r$
continues to be an affine parameter on each generator. The light-cone
gauge does not fully specify the coordinate system, and the metric
depends on six arbitrary constants that parametrize the 
residual gauge freedom. In Secs.~\ref{sec:corotating} and
\ref{sec:hor_lock} I introduce the corotating frame of the deformed
black hole, show that the precise definition of this frame determines
the six constants of the light-cone gauge, and express the metric in
the corotating coordinates $(v,r,\vartheta,\varphi)$, in which
$\vartheta = \theta$ and $\varphi = \phi - \omega_{\rm H} v$, with 
$\omega_{\rm H} := \chi_1/(4m_1)$ denoting the black hole's angular
velocity. The deformed event horizon admits a particularly simple
description in the corotating coordinates: its radial position stays
at $r=2m_1$, and its null generators move with constant values of
$\vartheta$ and $\varphi$. This allows me, in Sec.~\ref{sec:horizon},
to provide a simple and transparent description of the horizon's
intrinsic geometry.  

As I show in Sec.~\ref{sec:horizon}, the intrinsic geometry of a
slowly rotating, tidally deformed event horizon is captured by the
following expression for the Ricci scalar ${\cal R}$ of a
two-dimensional cross-section of the horizon,   
\begin{equation}   
(2m_1)^2 {\cal R} = 2 
- 8m_1^2 \E_{ab}(v)\, \Omega^a_\dagger \Omega^b_\dagger 
+ \frac{40}{3} m_1^2 \chi_{1\langle a}\B_{bc \rangle}(v)\,  
\Omega^a \Omega^b \Omega^c,
\label{hor_int_geom}
\end{equation} 
where angular brackets indicate the symmetric-tracefree operation --- 
completely symmetrize over indices and remove all traces --- and   
\begin{equation} 
\Omega^a := \bigl[ \sin\vartheta \cos(\varphi + \omega_{\rm H} v),
\sin\vartheta \sin(\varphi + \omega_{\rm H} v), \cos\vartheta \bigr] 
\end{equation} 
specifies the direction to a given point on the horizon, while 
\begin{equation} 
\Omega^a_\dagger := \bigl[ 
\sin\vartheta \cos(\varphi + \omega_{\rm H} v 
+ {\textstyle \frac{2}{3}} \chi_1), 
\sin\vartheta \sin(\varphi + \omega_{\rm H} v
+ {\textstyle \frac{2}{3}} \chi_1), 
\cos\vartheta \bigr] 
\end{equation} 
is a shifted version of the position vector. Except for this azimuthal 
shift, the term proportional to $\E_{ab}$ in Eq.~(\ref{hor_int_geom})
is identical to the one describing the tidal deformation of a
nonrotating black hole. The shift is a manifestation of the dragging
of inertial frames that accompanies the black hole's rotation. This
effect was first identified by Hartle in his 1974 seminal article
\cite{hartle:74}, but it should be noted that Hartle's calculation
contains a sign error that was previously documented by Fang and
Lovelace \cite{fang-lovelace:05}; Hartle obtains a shift of
$\frac{5}{3} \chi$ instead of $\frac{2}{3} \chi$. The dragging of
inertial frames also produces the octupolar deformation proportional
to $\chi_{1\langle a}\B_{bc \rangle}$; to the best of my knowledge,
this octupole component was never noticed before. My calculation,
carried out to first order in $\chi_1$, cannot reveal the azimuthal
shift of the octupole terms.  

A concrete description of the tidal deformation of a slowly rotating 
black hole requires the determination of the tidal moments
$\E_{ab}(v)$ and $\B_{ab}(v)$, and this task is undertaken in the
second part of this article. An important point is that the tidal
moments are determined as functions of advanced time $v$ in the local
rest frame of the black hole, and that they do not make a direct
reference to the spatial positions of the external matter. To
illustrate this point, and to provide a concrete example of
Eq.~(\ref{hor_int_geom}) as a description of tidal deformation on a
slowly rotating black hole, I consider a tidal environment in a state
of rigid rotation of angular frequency $\bar{\omega}$ around the black
hole's rotation axis. This could be produced, for example, by a
companion body on a circular orbit in the black hole's equatorial
plane. The scaling relations introduced previously imply that 
$m_1 \bar{\omega} \sim m_1/\tau \sim u^3 \ll u^2 \ll \chi_1$, which
means that $\bar{\omega}$ is necessarily much smaller than
$\omega_{\rm H}$. For such situations, the nonvanishing components of
$\E_{ab}$ and $\B_{ab}$ can be shown to be given by [refer to
Eqs.~(\ref{Ebar_components}) and (\ref{Bbar_components}) below]    
\begin{equation} 
\E_{11} = \E_0 + \E_2 \cos 2\bar{\phi}, \qquad 
\E_{12} = \E_2 \sin 2\bar{\phi}, \qquad 
\E_{22} = \E_0 - \E_2 \cos 2\bar{\phi}, \qquad 
\E_{33} = -2\E_0 
\label{E_components} 
\end{equation} 
and 
\begin{equation} 
\B_{13} = \B_1 \cos \bar{\phi}, \qquad 
\B_{23} = \B_1 \sin \bar{\phi}, 
\label{B_components} 
\end{equation} 
where $\E_0$, $\E_2$, and $\B_1$ are constants, and
$\bar{\phi} = \bar{\omega}(v-v_0)$ is the phase of the tidal field,
with $v_0$ an arbitrary constant that specifies the phase at $v=0$. As
we see from the displayed expressions, the tidal moments depend on $v$
through the phase $\bar{\phi}$. In a Newtonian context $\bar{\phi}$
could be identified with the azimuthal position of the companion body,
but no such interpretation is directly available in general
relativity. While $\bar{\phi}$ could still be related to the 
position of the companion body, this would require a mapping between
the body's position and a corresponding position on the event
horizon. A number of mappings were examined by Fang and Lovelace 
\cite{fang-lovelace:05} (see also Ref.~\cite{o'sullivan-hughes:14}), 
with the general conclusion that these constructions bring
arbitrariness and ambiguity in the description of the tidal
deformation. The main point that I wish to make is that a relation
between $\bar{\phi}$ and the position of the companion body is not
required; there is no arbitrariness nor ambiguity in the description
of the tidal deformation when $\bar{\phi}$ is properly viewed as a
phase function instead of an azimuthal position.      

Making the substitutions, we find that the deformation terms entering 
Eq.~(\ref{hor_int_geom}) are given by  
\begin{equation} 
\E_{ab}\, \Omega^a_\dagger \Omega^b_\dagger 
= \E_0 (1 - 3\cos^2\vartheta) 
+ \E_2 \sin^2\vartheta \cos 2 \bigl( \varphi 
+ \omega_{\rm H} v + {\textstyle \frac{2}{3}} \chi_1 - \bar{\phi} \bigr) 
\label{quad_deform} 
\end{equation} 
and 
\begin{equation} 
\chi_{1\langle a}\B_{bc \rangle}\, \Omega^a \Omega^b \Omega^c 
= -\frac{2}{5} \chi_1 \B_1 \sin\vartheta \bigl( 1 
- 5\cos^2\vartheta \bigr) \cos( \varphi + \omega_{\rm H} v 
- \bar{\phi}).   
\label{oct_deform} 
\end{equation} 
Equation (\ref{quad_deform}) indicates that the azimuthal position of
the quadrupole bulge is given by    
\begin{equation} 
\varphi_{\rm bulge} = (\bar{\phi} - \omega_{\rm H} v) - \frac{2}{3} \chi_1;  
\end{equation} 
the combination $\bar{\phi} - \omega_{\rm H} v$ is the phase of the
tidal field as measured in the black hole's corotating frame (in which
the horizon's generators are stationary), and $-\frac{2}{3}\chi_1$ is a
phase lag produced by the dragging of inertial frames. It is
interesting to note that in the Newtonian theory of tidal
interactions, a condition $\bar{\omega} \ll \omega_{\rm H}$ would
imply a tidal lead instead of a lag; as Hartle first observed
\cite{hartle:74}, a rotating black hole does not respect the Newtonian
relation, in spite of the fact that the condition 
$\bar{\omega}  \ll \omega_{\rm H}$ still leads to a decrease of the  
black hole's spin (see, for example, Sec.~IV D of
Ref.~\cite{poisson:04d}). On the other hand, Eq.~(\ref{oct_deform})
indicates that the octupole bulge is directly in phase with the tidal
field; an eventual phase shift could only be revealed in a calculation
carried out to second order in $\chi_1$. 

\subsection*{Determination of the tidal moments through $1.5\PN$
  order}   

My goal in the second part of this paper is to determine the tidal
moments $\E_{ab}$ and $\B_{ab}$ that characterize the geometry of a
slowly rotating, tidally deformed black hole. This requires immersing
the black hole in a patch of spacetime that extends well beyond the
domain $\N$, and specifying the conditions in this larger spacetime. I
assume that the mutual gravity between the black hole and the external
matter is sufficiently weak that it can be adequately represented as a
post-Newtonian ($\PN$) expansion. In terms of the scaling quantities
introduced previously, I assume that $u^2 \sim m/b \ll 1$; the black
hole, of course, it still taken to have strong internal gravity. I
therefore aim to obtain the tidal moments as post-Newtonian
expansions, with the specific goal to compute $\E_{ab}$ through order
$u^3$ beyond the Newtonian expression of order $m_2/b^3$, and to
compute $\B_{ab}$ through order $u$ beyond its leading-order
expression of order $m_2 u/b^3$. Overall this shall give us $1.5\PN$ 
accuracy for the tidal moments.  

\begin{figure} 
\includegraphics[width=0.5\linewidth]{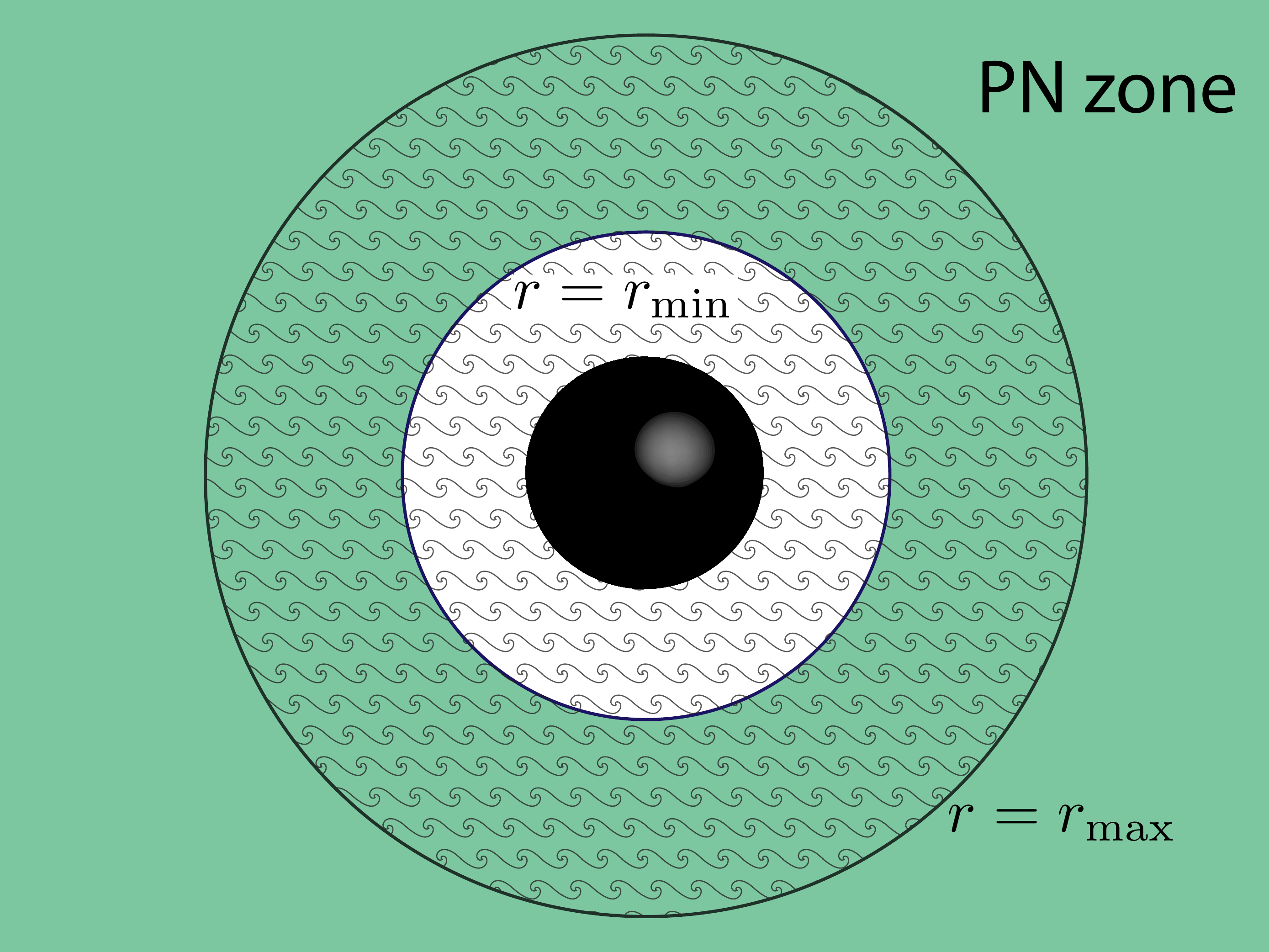}
\caption{The domain $\N$, shown in a wavy pattern, is bounded by 
$r < r_{\rm max} \ll b$. The post-Newtonian zone, shaded (shown in
green online), is bounded by $r > r_{\rm min} \gg m_1$. When 
$m_1 \ll b$ there is an overlap $r_{\rm min} < r < r_{\rm max}$ 
between $\N$ and the post-Newtonian zone.}    
\label{fig:f2} 
\end{figure} 

The calculation is based on the asymptotic matching of two metrics in
a common region of validity (see Fig.~\ref{fig:f2}). We have already
encountered the black-hole metric, which is restricted to the domain
$\N$ bounded by $r < r_{\rm max} \ll b$. I now introduce a
post-Newtonian metric to describe the mutual gravity between the black
hole and the external matter. This metric is restricted to a
post-Newtonian zone bounded by $r > r_{\rm min} \gg m_1$, beyond which
the black hole's gravity is too strong to be adequately described by
a post-Newtonian expansion. (The post-Newtonian zone is also bounded
externally by $r < \lambda$, with $\lambda$ a length scale for the
wavelength of the gravitational waves emitted by the entire system;
this is the boundary of the near zone, which plays no role in the
following developments.) When $m_1 \ll b$, the domain $\N$ and the  
post-Newtonian zone overlap when $r_{\rm min} < r < r_{\rm max}$, and
in this overlap both metrics are solutions to the Einstein field
equations that describe the same gravitational field; the metrics
must therefore agree once they are cast in the same coordinate
system. This observation delivers the tidal moments: Matching the
metrics in the overlap determines $\E_{ab}$ and $\B_{ab}$ in terms of
information provided by the post-Newtonian metric. Information flows
in the other direction as well: Because the post-Newtonian zone is
truncated at $r = r_{\rm min}$, the post-Newtonian metric contains
unknown functions that must be determined in terms of information
provided by the black-hole metric.   

The asymptotic matching of a black-hole metric to a post-Newtonian
metric was performed before \cite{yunes-etal:06,
  johnsonmcdaniel-etal:09, gallouin-etal:12, mundim-etal:14}, with the
specific objective of constructing initial data for the numerical
simulation of a binary black hole inspiral
\cite{reifenberger-tichy:12}. These works neglected the spin of each
black hole, except for Galloin, Nakano, Yunes, and Campanelli 
\cite{gallouin-etal:12}, who calculated the tidal moments of a
spinning black hole through $1\PN$ order. The calculation 
presented here is an improvement on this work, but I make no attempt 
to construct a global metric that could be used as initial data for
numerical evolutions. My interest is entirely on the tidal moments,
and I calculate them by exploiting the methods introduced by Taylor
and Poisson \cite{taylor-poisson:08}, who computed $\E_{ab}$ and
$\B_{ab}$ through $1\PN$ order for a nonspinning black hole.  

I begin in Sec.~\ref{sec:harmonic} by transforming the black-hole
metric from the light-cone coordinates $(v,r,\theta,\phi)$ to local
harmonic coordinates $(\bar{t},\bar{x}^a)$ that facilitate the
matching to the post-Newtonian metric; this system of harmonic
coordinates defines what I call the black-hole frame. In
Sec.~\ref{sec:tidal} I construct the post-Newtonian metric that
describes the mutual gravity between the black hole and the external
matter. The metric is first presented in global harmonic coordinates 
$(t,x^a)$ that define the post-Newtonian barycentric frame, and
expressed in terms of a Newtonian potential $U$, a vector potential
$U^a$, a post-Newtonian potential $\psi$, and a superpotential
$X$. It is next transformed to the black-hole frame and matched to the
black-hole metric. The matching determines $\E_{ab}$ and $\B_{ab}$,
which are given in terms of the potentials $U_{\rm ext}$, 
$U^a_{\rm ext}$, $\psi_{\rm ext}$, and $X_{\rm ext}$ that are sourced  
by the external matter. In Sec.~\ref{sec:2body} the tidal moments are
evaluated for the specific situation in which the black hole is a
member of a spinning two-body system, and in Sec.~\ref{sec:circular}
these results are further specialized to circular orbital motion (with
the spins either aligned or anti-aligned with the orbital angular
momentum). Finally, in Sec.~\ref{sec:Edot} I complete the calculation
of the tidal moments though $1.5\PN$ order by incorporating previously
neglected terms in the black-hole metric that involve $\dot{\E}_{ab}$,
the time derivative of the electric-type tidal moments. 

With the notation introduced in Eqs.~(\ref{E_components}) and
(\ref{B_components}), the tidal moments of a slowly rotating black
hole of mass $m_1$ and dimensionless spin $\chi_1$ in circular motion
around a companion body of mass $m_2$ and dimensionless spin $\chi_2$
are given by    
\begin{subequations} 
\label{EB_components} 
\begin{align}   
\E_0 &= -\frac{m_2}{2b^3} \biggl[ 1 
+ \frac{m_1}{2m} u^2 - 6 \frac{m_2}{m} \chi_2 u^3 
+ O(u^4) \biggr], \\ 
\E_2 &= -\frac{3m_2}{2b^3} \biggl[ 1 
- \frac{3m_1+4m_2}{2m} u^2 - 2 \frac{m_2}{m} \chi_2 u^3 
+ O(u^4) \biggr], \\ 
\B_1 &= -\frac{3m_2}{b^3} u \biggl[ 
1 - \frac{m_2}{m} \chi_2 u + O(u^2) \biggr], 
\end{align} 
\end{subequations} 
where $b$ is the orbital radius in harmonic coordinates, and $u$ 
is the orbital velocity, related to $b$ by the post-Newtonian relation 
\begin{equation} 
m/b = u^2 \bigl[ 1 + (3-\eta) u^2 
+ \tilde{\chi} u^3  + O(u^4) \bigr],   
\label{v2} 
\end{equation} 
with $m := m_1 + m_2$ denoting the total mass, $\eta := m_1 m_2/m^2$
the symmetric mass ratio, and 
\begin{equation} 
\tilde{\chi} := \frac{m_1(2m_1+3m_2)}{m^2}\, \chi_1
+ \frac{m_2(2m_2+3m_1)}{m^2}\, \chi_2. 
\end{equation} 
The phase of the tidal moments is given by 
\begin{equation} 
\bar{\phi} = \bar{\omega} (v-v_0) 
+ \frac{8}{3} \frac{m_1}{m} u^3 + O(u^4), 
\label{tidal_phase} 
\end{equation} 
where  
\begin{equation} 
\bar{\omega} = \sqrt{\frac{m}{b^3}} \biggl[ 1 
- \frac{1}{2} (3+\eta) u^2 - \frac{1}{2} \bar{\chi} u^3 
+ O(u^4) \biggr] 
\label{tidal_frequency} 
\end{equation} 
is the tidal angular frequency, expressed in terms of the
mass-weighted averaged spin  
\begin{equation} 
\bar{\chi} := \frac{m_1}{m^2} ( 2 m_1 + m_2 ) \chi_1
+ 3 \eta \chi_2. 
\label{chibar_final} 
\end{equation} 
The tidal frequency $\bar{\omega}$ is distinct from the orbital
frequency $\omega := u/b$, because $\omega$ is defined in terms of
time $t$ in the post-Newtonian barycentric frame, while $\bar{\omega}$
is defined in terms of advanced-time $v$ in the black hole's moving
frame. The frequencies also differ because the black-hole frame is slowly
precessing relative to the barycentric frame.

Equations (\ref{EB_components}), (\ref{tidal_phase}),
and (\ref{tidal_frequency}) are the final outcomes of the work carried
out in the second part of this article. While the $1.5\PN$ expressions
for the tidal quadrupole moments and tidal frequency are derived under 
the assumption that the black hole is rotating slowly 
($\chi_1 \ll 1$), it is important to observe that terms at $1.5\PN$
order in post-Newtonian expansions of the metric and equations of
motion are known to be linear in the spins. The $1.5\PN$ expressions,
therefore, are {\it valid to all orders in the spins}.   

The methods exploited here can be extended to provide a description of
the tidal deformation of a slowly rotating material body such as a
neutron star. The background metric outside such a body is still the  
one examined in Sec.~\ref{sec:background}, but the perturbation
constructed in Sec.~\ref{sec:perturbed} must be generalized to account 
for the body's gravitational Love numbers $k_2^{\rm el}$ and 
$k_2^{\rm mag}$ \cite{damour-nagar:09,
  binnington-poisson:09}. Additional Love numbers may be required to
describe the external geometry of the deformed body, and the metric 
inside the body must also be obtained and joined with the external
metric at the body's surface. These details are unlikely to influence
the results obtained in the second part of the paper: The tidal 
moments of Eqs.~(\ref{EB_components}) are expected to apply to both
black holes and material bodies. Work toward this generalization is
now underway, and will be reported in a forthcoming 
publication.\footnote{P. Landry and E. Poisson, in preparation.}    

To conclude, I note that the expressions of
Eqs.~(\ref{EB_components}) and (\ref{tidal_frequency}) are ready to be
inserted into the flux formulae obtained in
Ref.~\cite{chatziioannou-poisson-yunes:13} to describe the $1.5\PN$
tidal heating and torquing of a black hole, whether it is slowly or 
rapidly rotating. While expressions that were purported to be accurate
through $1.5\PN$ order were already presented in this work, these were
in fact incomplete, because they did not incorporate the spin terms in
the tidal moments, nor the spin terms in the tidal frequency. This 
situation will be remedied in a separate
publication.\footnote{K. Chatziioannou, E. Poisson, and N. Yunes, 
  in preparation.}      

\section{Irreducible potentials} 
\label{sec:potentials} 

In this section we construct the irreducible potentials that will 
appear in the metric of a slowly rotating, tidally deformed black 
hole. The construction is based on the methods introduced in Poisson
and Vlasov \cite{poisson-vlasov:10}, hereafter referred to as PV. In
this and the following sections, the black-hole mass will be denoted
$M$ instead of $m_1$, and its dimensionless spin will be denoted
$\chi^a$ instead of $\chi^a_1$. The notation employed in
Sec.~\ref{sec:intro} will be resumed in Sec.~\ref{sec:2body}, when the
black hole specifically becomes a member of a two-body system.   

The most primitive ingredients featured in the potentials are the
dimensionless spin pseudovector $\chi_a$, the electric-type tidal
quadrupole moment $\E_{ab}$, and the magnetic-type tidal quadrupole
moment $\B_{ab}$. These quantities are defined in the rest frame of
the black hole, in a local asymptotic region in which the
gravitational field is dominated by the tidal field of the external
spacetime. In this region the black hole's own gravitational field can
be neglected, and the black hole can be viewed as a test body moving
on a (potentially accelerated) world line $\gamma$; this test body
possesses a mass $M$ and a spin tensor $S^{\mu\nu}$ as measured in the 
local asymptotic rest frame. 

The definitions of $\chi_a$, $\E_{ab}$, and $\B_{ab}$ rely on a triad
of vectors $e^\alpha_a$ on $\gamma$; these are mutually orthonormal, 
orthogonal to $\gamma$'s tangent vector $u^\alpha$ (the black hole's
velocity vector), and Fermi-Walker transported on the world line. The 
dimensionless spin pseudovector is defined by $\chi_a := S_a/M^2$,
where $S_a := \frac{1}{2} \epsilon_{apq} e^p_\mu e^q_\nu S^{\mu\nu}$
is the spin pseudovector, $\epsilon_{apq}$ is the completely
antisymmetric permutation symbol, and latin indices are lowered and
raised freely with the Euclidean metric. The dimensionless spin
$\chi_a$ is taken to be numerically small, to reflect the assumption
that the black hole is rotating slowly. The tidal quadrupole moments 
are defined in terms of the Weyl tensor of the external spacetime
evaluated on $\gamma$; we have 
$\E_{ab} := C_{a0b0}$ and 
$\B_{ab} := \frac{1}{2} \epsilon_{apq} C^{pq}_{\ \ b0}$, where  
$C_{a0b0} := C_{\alpha\mu\beta\nu} e^\alpha_a
u^\mu e^\beta_b u^\nu$ and $C_{abc0} := C_{\alpha\beta\gamma\mu}
e^\alpha_a e^\beta_b e^\gamma_c u^\mu$. By virtue of the
symmetries of the Weyl tensor, $\E_{ab}$ and $\B_{ab}$ are both
symmetric and tracefree. 

In the context of this work, a parity transformation is a reflection
of the triad described by $e^\alpha_a \to -e^\alpha_a$; the transformation
keeps $\epsilon_{abc}$ unchanged. Under a parity transformation the
spin pseudovector and tidal moments change according to 
\begin{equation} 
\chi_a \to \chi_a, \qquad 
\E_{ab} \to \E_{ab}, \qquad 
\B_{ab} \to -\B_{ab}. 
\end{equation} 
We observe that while $\E_{ab}$ transforms as an ordinary Cartesian
tensor under a parity transformation, $\B_{ab}$ transforms as a
pseudotensor; $\chi_a$ transforms as a pseudovector, as expected of a
quantity describing spin. We shall say that $\E_{ab}$ has even parity,
while $\chi_a$ and $\B_{ab}$ have odd parity. 

The potentials are constructed by following the rules described in
Sec.~II and Appendix A of PV. We consider first a construction
involving Cartesian coordinates $x^a$, and describe next a 
construction involving a related system of spherical polar coordinates
$(r,\theta^A)$, in which $\theta^A = (\theta,\phi)$. The relation
between the two systems is the usual $x^a = r \Omega^a(\theta^A)$, in
which $\Omega^a$ is the unit vector $x^a/r$ expressed in terms of the
angles $\theta^A$; explicitly $\Omega^a = (\sin\theta \cos\phi,
\sin\theta \sin\phi, \cos\theta)$. 

The potentials are obtained by combining $\chi_a$, $\E_{ab}$,
$\B_{ab}$, and $\Omega^a$ in various irreducible ways, with each
potential carrying a specific multipole order $\ell$ and a specific
parity label (even or odd). For our purposes here (see
Sec.~\ref{sec:intro}) we construct potentials that are linear in
$\chi_a$, $\E_{ab}$, $\B_{ab}$, as well as bilinear potentials that
couple $\chi_a$ to $\E_{ab}$ and $\chi_a$ to $\B_{ab}$. The nonlinear
couplings between $\E_{ab}$ and $\B_{ab}$ were considered by PV, and
we neglect couplings between $\chi^a$ and itself --- all expressions
will be linearized with respect to the dimensionless spin.   

Because $\chi_a$ is of odd parity, it is to be involved in the
construction of an odd-parity rotational potential. Following the
general prescription of Eq.~(A9) of PV, we introduce
\begin{equation} 
\chid_a := \epsilon_{abc} \Omega^b \chi^c; 
\label{chid_def} 
\end{equation} 
this is a vectorial, dipolar $(\ell=1)$ potential, as indicated by the 
label $\sf d$. The tidal potentials associated with $\E_{ab}$ and
$\B_{ab}$ are constructed in Sec.~II of PV. We have 
\begin{equation} 
\Eq := \E_{ab} \Omega^a \Omega^b, \qquad 
\Eq_a := \gamma_a^{\ e} \E_{ec} \Omega^c, \qquad 
\Eq_{ab} := 2 \gamma_a^{\ e} \gamma_b^{\ f} \E_{ef} 
+ \gamma_{ab} \Eq,
\label{Eq_def} 
\end{equation} 
where $\gamma_a^{\ b} := \delta_a^{\ b} - \Omega_a \Omega^b$ is a
projector to the subspace transverse to $\Omega_a$; these are
even-parity potentials of quadrupole order $(\ell = 2)$, as indicated
by the label $\sf q$. We also have 
\begin{equation} 
\Bq_a := \epsilon_{aef} \Omega^e \B^f_{\ c} \Omega^c, \qquad 
\Bq_{ab} := \epsilon_{aef} \Omega^e \B^f_{\ c} \gamma^c_{\ b}
+ \epsilon_{bef} \Omega^e \B^f_{\ c} \gamma^c_{\ a}, 
\label{Bq_def}
\end{equation} 
and these are odd-parity potentials, also of quadrupole order.  

The coupling of $\chi_a$ and $\E_{ab}$ produces the pseudotensors 
\begin{equation} 
\F_a := \E_{ab} \chi^b, \qquad 
\F_{abc} := \E_{\langle a b} \chi_{c\rangle}, 
\label{F_def} 
\end{equation}  
in which the angular brackets designate the operation of
symmetrization and trace removal, so that $\F_{abc}$ is a
symmetric-tracefree tensor. These give rise to the dipolar $(\ell=1)$,
odd-parity potential  
\begin{equation} 
\Fd_a := \epsilon_{abc} \Omega^b \F^c
\label{Fd_def}
\end{equation} 
and the octupolar ($\ell=3$), odd-parity potentials 
\begin{equation} 
\Fo_a := \epsilon_{aef} \Omega^e \F^f_{\ rs} \Omega^r \Omega^s, \qquad 
\Fo_{ab} := \bigl( \epsilon_{aef} \Omega^e \F^f_{\ cr} \gamma^c_{\ b}  
+ \epsilon_{bef} \Omega^e \F^f_{\ cr} \gamma^c_{\ a} \bigr) \Omega^r. 
\label{Fo_def}
\end{equation} 
On the other hand, the coupling of $\chi_a$ and $\B_{ab}$ produces the
tensors 
\begin{equation} 
\K_a := \B_{ab} \chi^b, \qquad 
\K_{abc} := \B_{\langle a b} \chi_{c\rangle}, 
\label{K_def} 
\end{equation} 
the dipolar, even-parity potentials 
\begin{equation} 
\Kd := \K_a \Omega^a, \qquad 
\Kd_a := \gamma_a^{\ e} \K_e,
\label{Kd_def}
\end{equation} 
and the octupolar, even-parity potentials 
\begin{equation} 
\Ko := \K_{abc} \Omega^a \Omega^b \Omega^c, \qquad 
\Ko_a := \gamma_a^{\ e} \K_{ebc} \Omega^b \Omega^c, \qquad 
\Ko_{ab} := 2 \gamma_a^{\ e} \gamma_b^{\ f} \K_{efc} \Omega^c  
+ \gamma_{ab} \Ko. 
\label{Ko_def}
\end{equation} 

We next convert the Cartesian potentials introduced previously into
angular-coordinate potentials. The relations 
$x^a = r\Omega^a(\theta^A)$ give rise to the transformation matrix 
$\Omega^a_A := \partial \Omega^a/\partial \theta^A$, and the
transformation of the potentials is described by 
\begin{equation} 
\Eq_A := \Eq_a \Omega^a_A, \qquad 
\Eq_{AB} := \Eq_{ab} \Omega^a_A \Omega^a_B,
\end{equation} 
with similar relations defining $\chid_A$, $\Fo_{AB}$, and so on. 

It is helpful to decompose the angular-coordinate potentials in
scalar, vector, and tensor spherical harmonics. The methods to achieve
this are described in Sec.~II and Appendix A of PV. The
decomposition involves the even-parity harmonics 
\begin{equation} 
Y_A^{\ell\m} := D_A Y^{\ell\m}, \qquad 
Y_{AB}^{\ell\m} := \Bigl[ D_A D_B + \frac{1}{2} \ell(\ell+1)
\Omega_{AB} \Bigr] Y^{\ell\m} 
\label{even_harm} 
\end{equation} 
and the odd-parity harmonics 
\begin{equation} 
X_A^{\ell\m} := -\epsilon_A^{\ B} D_B Y^{\ell\m}, \qquad 
X_{AB}^{\ell,\m} := -\frac{1}{2} \bigl( \epsilon_A^{\ C} D_B 
+ \epsilon_B^{\ C} D_A \bigr) D_C Y^{\ell\m}. 
\label{odd_harm} 
\end{equation}  
Here $\Omega_{AB} = \mbox{diag}(1,\sin^2\theta)$ is the metric on a
unit two-sphere, and $D_A$ is the covariant-derivative operator
compatible with this metric; $\epsilon_{AB}$ is the Levi-Civita tensor
on the unit two-sphere ($\epsilon_{\theta\phi} = \sin\theta$), and its
index is raised with $\Omega^{AB}$, the matrix inverse to
$\Omega_{AB}$. It should be noted that the tensorial harmonics are
tracefree, in the sense that $\Omega^{AB} Y^{\ell\m}_{AB} = 
\Omega^{AB} X^{\ell\m}_{AB} = 0$. It will be a consistent convention
in this work that tensorial operations on spherical harmonics refer to
the unit two-sphere (instead of a sphere of radius $r$).  

The starting point for the decomposition are the identities  
\begin{subequations}
\label{sph_harm_decomp1}  
\begin{align} 
& \chi_a \Omega^a =\sum_\m \chid_\m Y^{1\m}, \qquad
\F_{a} \Omega^a = \sum_\m \Fd_\m Y^{1\m}, \qquad    
\K_{a} \Omega^a = \sum_\m \Kd_\m Y^{1\m}, \\     
& \E_{ab} \Omega^a \Omega^b = \sum_\m \Eq_\m Y^{2\m}, \qquad  
\B_{ab} \Omega^a \Omega^b = \sum_\m \Bq_\m Y^{2\m}, \\ 
& \F_{abc} \Omega^a \Omega^b \Omega^c 
= \sum_\m \Fo_\m Y^{3\m}, \qquad       
\K_{abc} \Omega^a \Omega^b \Omega^c 
= \sum_\m \Ko_\m Y^{3\m},  
\end{align} 
\end{subequations} 
which involve the (scalar) spherical-harmonic functions
$Y^{\ell\m}(\theta^A)$. The identities provide a packaging of the 3
independent components of $\chi_a$, $\F_a$, and $\K_a$ in the
coefficients $\chid_\m$, $\Fd_\m$, and $\Fo_\m$, a packaging of the
5 independent components of $\E_{ab}$ and $\B_{ab}$ in $\Eq_\m$ and
$\Bq_\m$, and a packaging of the 7 independent components of
$\F_{abc}$ and $\K_{abc}$ in $\Fo_\m$ and $\Ko_\m$. The precise
definition of the various coefficients in Eq.~(\ref{sph_harm_decomp1})
depend on the conventions adopted for the spherical harmonics. These
are spelled out in Table~\ref{tab:Ylm}, and the coefficients are
defined in Table~\ref{tab:coeffs}. To simplify all expressions we
align $\chi_a$ with the polar axis, so that $\chi_a = (0,0,\chi)$.

%\begingroup
%\squeezetable
\begin{table}
\caption{Spherical-harmonic functions $Y^{\ell\m}$. The functions are
  real, and they are listed for the relevant modes $\ell=1$ (dipole),
  $\ell=2$ (quadrupole), and $\ell=3$ (octupole). The abstract index
  $\m$ describes the dependence of these functions on the angle
  $\phi$; for example $Y^{\ell,2s}$ is proportional to $\sin2\phi$. To
  simplify the expressions we write $C := \cos\theta$ and 
  $S := \sin\theta$.} 
\begin{ruledtabular} 
\begin{tabular}{l} 
$ Y^{1,0} = C $ \\ 
$ Y^{1,1c} = S \cos\phi $ \\ 
$ Y^{1,1s} = S \sin\phi $ \\ 
\\ 
$ Y^{2,0} = 1-3C^2 $ \\ 
$ Y^{2,1c} = 2SC\cos\phi $ \\ 
$ Y^{2,1s} = 2SC\sin\phi $ \\ 
$ Y^{2,2c} = S^2\cos 2\phi $ \\ 
$ Y^{2,2s} = S^2\sin 2\phi $ \\ 
\\
$ Y^{3,0} = C(3-5C^2) $ \\  
$ Y^{3,1c} = \frac{3}{2} S(1-5C^2)\cos\phi $ \\  
$ Y^{3,1s} = \frac{3}{2} S(1-5C^2)\sin\phi $ \\  
$ Y^{3,2c} = 3S^2 C \cos 2\phi $ \\  
$ Y^{3,2s} = 3S^2 C \sin 2\phi $ \\  
$ Y^{3,3c} = S^3 \cos 3\phi $ \\  
$ Y^{3,3s} = S^3 \sin 3\phi $
\end{tabular}
\end{ruledtabular} 
\label{tab:Ylm} 
\end{table} 
%\endgroup

%\begingroup
%\squeezetable
\begin{table}
\caption{Spherical-harmonic coefficients of
  Eq.~(\ref{sph_harm_decomp1}). The relations between $\Kd_\m$,  
  $\K_a$, and $\chi \Bq_\m$ are identical to those between $\Fd_\m$,   
  $\F_a$, and $\chi \Eq_\m$. Similarly, the relations between
  $\Ko_\m$, $\K_{abc}$, and $\chi \Bq_\m$ are identical to those
  between $\Fo_\m$, $\F_{abc}$, and $\chi \Eq_\m$.}  
\begin{ruledtabular} 
\begin{tabular}{l} 
$ \chid_0 = \chi_3 = \chi$ \\ 
$ \chi_{1c} = \chi_1 = 0$ \\ 
$ \chi_{1s} = \chi_2 = 0$ \\ 
\\ 
$ \Eq_0 = \frac{1}{2} (\E_{11} + \E_{22}) $ \\ 
$ \Eq_{1c} = \E_{13} $ \\ 
$ \Eq_{1s} = \E_{23} $ \\ 
$ \Eq_{2c} = \frac{1}{2} ( \E_{11} - \E_{22}) $ \\ 
$ \Eq_{2s} = \E_{12} $ \\ 
\\ 
$ \Fd_0 = \F_3 = -2\chi \Eq_0 $ \\ 
$ \Fd_{1c} = \F_1 = \chi \Eq_{1c} $ \\ 
$ \Fd_{1s} = \F_2 = \chi \Eq_{1s} $ \\ 
\\ 
$ \Fo_0 = \frac{1}{2} (\F_{113} + \F_{223}) 
= \frac{3}{5} \chi \Eq_{0} $ \\ 
$ \Fo_{1c} = \frac{1}{2} (\F_{111} + \F_{122}) 
= -\frac{4}{15} \chi \Eq_{1c} $ \\  
$ \Fo_{1s} = \frac{1}{2} (\F_{112} + \F_{222}) 
= -\frac{4}{15} \chi \Eq_{1s} $ \\  
$ \Fo_{2c} = \frac{1}{2} (\F_{113} - \F_{223}) 
= \frac{1}{3} \chi \Eq_{2c} $ \\  
$ \Fo_{2s} = \F_{123} = \frac{1}{3} \chi \Eq_{2s} $ \\  
$ \Fo_{3c} = \frac{1}{4} (\F_{111} - 3\F_{122}) = 0 $ \\ 
$ \Fo_{2s} = \frac{1}{4} (3\F_{112} - F_{222}) = 0 $
\end{tabular}
\end{ruledtabular} 
\label{tab:coeffs} 
\end{table} 
%\endgroup

The decomposition of the potentials in spherical harmonics is
described by 
\begin{subequations} 
\label{sph_harm_decomp2}  
\begin{align}
& \chid_A = \sum_\m \chid_\m X^{1\m}_A, 
\label{sph_harm_decomp2a} \\
& \Fd_A = \sum_\m \Fd_\m X^{1\m}_A, \\
& \Kd = \sum_\m \Kd_\m Y^{1\m}, \qquad 
\Kd_A = \sum_\m \Kd_\m Y^{1\m}_A, \\ 
& \Eq = \sum_\m \Eq_\m Y^{2\m}, \qquad 
\Eq_A = \frac{1}{2} \sum_\m \Eq_\m Y^{2\m}_A, \qquad
\Eq_{AB} = \sum_\m \Eq_\m Y^{2\m}_{AB}, \\ 
& \Bq_A = \frac{1}{2} \sum_\m \Bq_\m X^{2\m}_A, \qquad 
\Bq_{AB} = \sum_\m \Bq_\m X^{2\m}_{AB}, \\ 
&  \Fo_A = \frac{1}{3} \sum_\m \Fo_\m X^{3\m}_A, \qquad 
\Fo_{AB} = \frac{1}{3} \sum_\m \Fo_\m X^{3\m}_{AB}, \\
& \Ko = \sum_\m \Ko_\m Y^{3\m}, \qquad 
\Ko_A = \frac{1}{3} \sum_\m \Ko_\m Y^{3\m}_A, \qquad 
\Ko_{AB} = \frac{1}{3} \sum_\m \Ko_\m Y^{3\m}_{AB}. 
\end{align} 
\end{subequations} 
It should be noted that tensorial potentials are not defined when
$\ell = 1$. The origin of the numerical coefficients that appear in
front of the sums is explained in Eqs.~(A7), (A8), (A14), and (A15) of
PV.  

\section{Background spacetime} 
\label{sec:background} 

The metric of a slowly rotating black hole can be obtained from the
exact Kerr metric by neglecting all terms beyond linear order in the
spin parameter $\chi := S/M^2$, in which $S$ is the black hole's
angular momentum and $M$ is its mass. In the original Boyer-Lindquist
coordinates $(t,r,\theta,\tilde{\phi})$, the metric is given by   
\begin{equation} 
ds^2 = -f\, dt^2 + f^{-1}\, dr^2 
+ r^2 \bigl( d\theta^2 + \sin^2\theta\, d\tilde{\phi}^2 \bigr) 
- 2\frac{2\chi M^2}{r} \sin^2\theta\, dt d\tilde{\phi},
\end{equation} 
where $f := 1-2M/r$. The Boyer-Lindquist coordinates are singular at
the event horizon, and we replace them with light-cone coordinates 
$(v,r,\theta,\phi)$ defined by 
\begin{equation} 
dv = dt + f^{-1}\, dr, \qquad 
d\phi = d\tilde{\phi} + \frac{2\chi M^2}{r^3 f}\, dr. 
\end{equation} 
The coordinate transformation produces 
\begin{equation} 
ds^2 = -f\, dv^2 + 2\, dvdr + r^2 d\Omega^2 
- 2\frac{2\chi M^2}{r} \sin^2\theta\, dv d\phi, 
\label{background_metric1} 
\end{equation} 
where $d\Omega^2 := \Omega_{AB} d\theta^A d\theta^B := d\theta^2 
+ \sin^2\theta\, d\phi^2$. The transformation from $\tilde{\phi}$ to
$\phi$ was introduced to remove a term proportional to 
$dr d\tilde{\phi}$ in the metric, produced when $dt$ is replaced by   
$dv = dt + f^{-1} dr$. 

The coordinates $(v,r,\theta,\phi)$ differ from the coordinates
$(\hat{v},r,\theta,\hat{\phi})$ attached to the incoming principal
null congruence of the Kerr spacetime (see, for example, Sec.~5.3.6 of
Ref.~\cite{poisson:b04}). While $\hat{v} = v + O(\chi^2)$, we have that 
$d\hat{\phi} = d\tilde{\phi} + (\chi M/r^2 f)\, dr + O(\chi^2) 
= d\phi + (\chi M/r^2)\, dr + O(\chi^2)$, so that 
\begin{equation} 
\hat{\phi} = \phi - \chi \frac{M}{r} + O(\chi^2);
\label{phi_hat} 
\end{equation} 
the constant of integration was selected to ensure that
$\hat{\phi} = \phi$ when $r=\infty$. 

The coordinates $(v,r,\theta,\phi)$ are tied to the behavior of
incoming null geodesics that are tangent to converging null cones. It
is easy to show that $\ell_\alpha = -\partial_\alpha v$ is null, so
that each surface $v = \mbox{constant}$ is a null
hypersurface. The vector $\ell^\alpha$ is tangent to its null
generators, and the only nonvanishing component is $l^r = -1$; this
indicates that $v$, $\theta$, and $\phi$ are all constant on the
generators, and that $-r$ is an affine parameter. We note that the
null geodesics have zero angular momentum, because the component of   
$\ell_\alpha$ along the azimuthal Killing vector $\phi^\alpha$
vanishes.  It can also be noted that $\tilde{\phi}$ increases on the
geodesics, while $\hat{\phi}$ decreases.    

In the notation introduced in Sec.~\ref{sec:potentials}, the
components of the background metric can be expressed as 
\begin{equation} 
g_{vv} = -f, \qquad 
g_{vr} = 1, \qquad 
g_{vA} = \frac{2M^2}{r} \chid_A, \qquad 
g_{AB} = r^2 \Omega_{AB}, 
\label{background_metric2}
\end{equation} 
where $\chid_A$ is the rotational potential of
Eq.~(\ref{sph_harm_decomp2a}). 

\section{Perturbed spacetime} 
\label{sec:perturbed} 

Our goal in this section is to add a tidal perturbation to the
background metric of Eq.~(\ref{background_metric2}). As explained in
Sec.~\ref{sec:intro}, we focus our attention on a domain $\N$ empty of
matter, bounded by $r < r_{\rm max} \ll b$, with $b$ denoting the
distance scale to the external matter. We follow the general methods
detailed in Poisson and Vlasov (\cite{poisson-vlasov:10}: PV). In
particular, we continue to work in light-cone coordinates, and insist
that the coordinates $(v,r,\theta,\phi)$ keep their geometrical
meaning in the perturbed spacetime: $v$ shall continue to be constant
on converging null cones, $\theta$ and $\phi$ shall continue to be
constant on each generator, and $-r$ shall continue to be an affine
parameter on each generator. As shown in Sec.~V A of PV, these
requirements imply that $g_{vr} = 1$, $g_{rr} = 0 = g_{rA}$, so that
$g_{vv}$, $g_{vr}$, $g_{vA}$, and $g_{AB}$ are the only nonvanishing
components of the metric. In addition, we shall see that the field
equations allow us to preserve the meaning of $r$ as an areal radius,
in the sense that $4\pi r^2$ measures the area of a two-surface of
constant $v$ and $r$.       

At a large distance from the black hole (where $r$ is such that 
$2M \ll r < r_{\rm max}$), the metric is dominated by the tidal field,
which is characterized by the tidal moments $\E_{ab}(v)$ and
$\B_{ab}(v)$. As shown in Eq.~(3.4) of PV, the asymptotic form of the 
metric is given by 
\begin{equation} 
g_{vv} \sim -r^2 \Eq, \qquad 
g_{vA} \sim -\frac{2}{3} r^3 \bigl( \Eq_A - \Bq_A \bigr), \qquad 
g_{AB} \sim -\frac{1}{3} r^4 \bigl( \Eq_{AB} - \Bq_{AB} \bigr), 
\label{metric_asympt} 
\end{equation} 
in addition to the exact statement $g_{vr} = 1$. These expressions are
accurate to leading order in an expansion in powers of $r/b$, and we
shall maintain this degree of accuracy throughout this work. When $r$
becomes comparable to $2M$ the information contained in
Eq.~(\ref{background_metric2}) becomes important, and what is required
is a perturbed metric that incorporates this information in addition
to the asymptotic behavior captured by Eq.~(\ref{metric_asympt}). As
indicated previously, we wish the metric to contain terms that are
linear in $\chi$, $\E_{ab}$, $\B_{ab}$, as well as terms that scale as
$\chi \E_{ab}$ and $\chi \B_{ab}$; we are satisfied with the neglect
of higher-order terms in $\chi$ and higher-order terms in the tidal
moments. 

To construct this metric it is helpful to alter our perspective and
take the background spacetime to have the metric 
\begin{equation} 
g^{\rm back}_{vv} = -f, \qquad 
g^{\rm back}_{vr} = 1, \qquad 
g^{\rm back}_{AB} = r^2 \Omega_{AB} 
\label{Schwarzschild} 
\end{equation} 
of a nonrotating black hole. In this new perspective, the missing
rotational term $p_{vA} = (2M^2/r) \chid_A$ is treated as a
perturbation, which can be added to the tidal perturbation  
\begin{equation} 
p_{vv} = -r^2 \eq_1 \Eq, \qquad 
p_{vA} = -\frac{2}{3} r^3 \bigl( \eq_4\, \Eq_A 
- \bq_4\, \Bq_A \bigr), \qquad 
p_{AB} = -\frac{1}{3} r^4 \bigl( \eq_7\, \Eq_{AB} - \bq_7\, \Bq_{AB} 
\bigr)
\end{equation} 
first constructed in Ref.~\cite{poisson:05} (see also
Refs.~\cite{detweiler:01, detweiler:05}). The perturbed metric is
expressed as $g_{\alpha\beta} =  g^{\rm back}_{\alpha\beta} 
+ p_{\alpha\beta}$, in which $g^{\rm back}_{\alpha\beta}$ is the new
background metric of Eq.~(\ref{Schwarzschild}), and $p_{\alpha\beta}$
is a sum of rotational and tidal perturbations. The radial
functions $\eq_n$ and $\bq_n$ are displayed
Table~\ref{tab:radial}. While the expressions given here incorporate
the leading rotational and tidal deformations, we wish to go beyond
the leading order and obtain improved expressions that capture the
couplings between the rotational and tidal terms. In the new
perspective adopted here, this requires us to go beyond first-order
perturbation theory.  

The expansions of Eqs.~(\ref{sph_harm_decomp2}) imply that the
first-order perturbation $p_{\alpha\beta}$ admits the usual
decomposition in spherical harmonics. In the even-parity sector we
have  
\begin{equation}
p_{vv} = \sum_{\ell\m} h_{vv}^{\ell\m}\, Y^{\ell\m}, \qquad
p_{vA} = \sum_{\ell\m} j_v^{\ell\m}\, Y^{\ell\m}_A, \qquad 
p_{AB} = r^2 \sum_{\ell\m} G^{\ell\m}\, Y^{\ell\m}_{AB},  
\label{even_sector} 
\end{equation}  
in which the coefficients $h^{\ell\m}_{vv}$, $j^{\ell\m}_v$, and  
$G^{\ell\m}$ depend on $r$ only. The decompositions, in fact, imply
that the even-parity perturbation is made up entirely of a tidal
deformation created by $\E_{ab}$, which is purely quadrupolar 
($\ell=2$). The decompositions omit a term $r^2 K^{2\m}\, \Omega_{AB} 
Y^{2\m}$ that could also be included in $p_{AB}$; this refinement of
the light-cone gauge is allowed by the field equations (refer to
Ref.~\cite{preston-poisson:06b}), and it ensures that $r$ continues to
be an areal radius in the perturbed spacetime. In the odd-parity
sector we have    
\begin{equation} 
p_{vv} = 0, \qquad 
p_{vA} = \sum_{\ell\m} h_v^{\ell\m}\, X^{\ell\m}_A, \qquad 
p_{AB} = \sum_{\ell\m} h_2^{\ell\m}\, X^{\ell\m}_{AB}, 
\label{odd_sector}
\end{equation}  
in which $h_v^{\ell\m}$ and $h_2^{\ell\m}$ depend on $r$ only. In this
case the decompositions imply that the odd-parity perturbation
consists of a dipolar $(\ell = 1)$ rotational perturbation and a
quadrupolar tidal deformation created by $\B_{ab}$.  

When the first-order perturbation is used as a seed for a second-order
calculation of the perturbed metric, the $\ell=1$ terms associated
with the rotational perturbation couple to the $\ell=2$ tidal
terms. The composition of the relevant spherical harmonics implies
that the resulting perturbation will have contributions at multipole
orders $\ell=1$, $\ell=2$, and $\ell=3$. The metric can then be
obtained by creating an ansatz that incorporates all possible
contributions at these multipole orders, substituting it into the
Einstein field equations, and solving for the unknown functions of
$r$. The ansatz must respect the parity rules implied by the
composition of spherical harmonics, and after some thought it becomes
clear that it will implicate the irreducible potentials introduced in
Sec.~\ref{sec:potentials}. Our metric ansatz is given by 
\begin{subequations} 
\label{metric} 
\begin{align} 
g_{vv} &= -f - r^2 \eq_1\, \Eq - r^2 \ehatq_1\, \chi \partial_\phi \Eq 
+ r^2 \kd_1\, \Kd - r^2 \ko_1\, \Ko, \\ 
g_{vr} &= 1, \\ 
g_{vA} &= \frac{2M^2}{r} \chid_A 
- \frac{2}{3} r^3 \bigl( \eq_4\, \Eq_A - \bq_4\, \Bq_A \bigr) 
- r^3\, \chi \partial_\phi \bigl( \ehatq_4 \Eq_A 
- \bhatq_4\, \Bq_A \bigr) 
- r^3 \bigl( \fd_4\, \Fd_A - \kd_4\, \Kd_A \bigr) 
+ r^3 \bigl( \fo_4\, \Fo_A + \ko_4\, \Ko_A \bigr), \\ 
g_{AB} &= r^2 \Omega_{AB} 
- \frac{1}{3} r^4 \bigl( \eq_7\, \Eq_{AB} - \bq_7\, \Bq_{AB} \bigr) 
- r^4\, \chi \partial_\phi \bigl( \ehatq_7\, \Eq_{AB} 
- \bhatq_7\, \Bq_{AB} \bigr)
- r^4 \bigl( \fo_7\, \Fo_{AB} - \ko_7\, \Ko_{AB} \bigr),
\end{align} 
\end{subequations} 
in which $\ehatq_n$, $\bhatq_n$, $\kd_n$, $\ko_n$, $\fd_n$, and
$\fo_n$ are functions of $r$ to be determined by solving the Einstein
field equations to second order in perturbation theory. Notice
that the ansatz omits terms in $p_{AB}$ that are proportional to
$\Omega_{AB}$; all the contributions are tracefree, in the
sense that $\Omega^{AB} p_{AB} = 0$. This feature was also observed in
the first-order perturbation, where it was identified as a refinement
of the light-cone gauge. We have verified that adding terms $r^2
K^{\ell\m}\, \Omega_{AB} Y^{\ell\m}$ to the metric produces
differential equations for $K^{\ell\m}(r)$ that are entirely decoupled
from the remaining field equations. The simplest solution is 
$K^{\ell\m} = 0$. The freedom to refine the light-cone gauge is
therefore preserved at second order, and $r$ continues to be an areal
radius in the metric of Eq.~(\ref{metric}).      

We note that expressions such as $\partial_\phi \Eq_A$ and 
$\partial_\phi \Bq_{AB}$ appear to break the covariance of the metric
with respect to transformations of the angular coordinates
$\theta^A$. This violation, however, is only apparent. An equivalent
covariant expression is easily obtained by introducing a vector
$\phi^A$ on each two-sphere $(v,r) = \mbox{constant}$ that generates a 
rotation around the axis identified by the spin pseudovector
$\chi_a$. In our canonical coordinates in which $\chi_a$ is aligned
with the polar axis, $\phi^A = (0,1)$, and 
\begin{equation} 
\partial_\phi \Eq_A = {\cal L}_\phi \Eq_A, \qquad 
\partial_\phi \Bq_{AB} = {\cal L}_\phi \Bq_{AB},
\label{Lie} 
\end{equation} 
where ${\cal L}_\phi$ denotes the Lie derivative in the direction of
the vector $\phi^A$. The right-hand side of each equality is covariant,
and $\partial_\phi$ can be freely replaced with ${\cal L}_\phi$ in the
metric ansatz of Eq.~(\ref{metric}).   

The metric of Eq.~(\ref{metric}) is used to calculate the Ricci tensor 
$R_{\alpha\beta}$, and time derivatives are neglected by virtue of our
assumption that the tidal moments vary slowly. The Ricci tensor is 
then expanded in powers of $\chi$, $\E_{ab}$, and $\B_{ab}$. The
expansion includes contributions that are linear in these quantities,
and these vanish when the appropriate substitutions are made for the
radial functions $\eq_n$ and $\bq_n$. The expansion also includes
terms that scale as $\chi \E_{ab}$ and $\chi \B_{ab}$, but it neglects
higher-order terms. The contributions to $R_{\alpha\beta}$
proportional to $\chi \E_{ab}$ and $\chi \B_{ab}$ are collected in a
spherical-harmonic decomposition, and the vacuum field equations imply 
that each coefficient must vanish separately. This calculation gives
rise to a number of differential equations for the remaining radial
functions. 

The solutions to these equations feature two types of integration
constants. As an example, let us consider the radial function $\ko_1$,
which appears in $g_{vv}$ as the coefficient in front of $\Ko$, a
bilinear potential coupling $\chi_a$ to $\B_{ab}$. The equation
satisfied by $\ko_1$ involves a linear differential operator inherited
from first-order perturbation theory, as well as a nonlinear source
term generated by the first-order perturbation. (The actual situation
is more complicated, because $\ko_1$ is implicated with other radial
functions in a system of coupled equations. But this complication has
no bearing on this discussion, which for clarity we choose to frame in
a simplified form.) The general solution to the equation is       
\begin{equation} 
\ko_1 = 2\frac{M^2}{r^2} - \frac{8}{3} \frac{M^3}{r^3} 
+ \cc{o} \frac{M^4}{r^4} 
+ d^{\scriptstyle \sf o}\frac{r}{M} f^2 \biggl(1 - \frac{M}{r} \biggr), 
\end{equation} 
and it depends on two arbitrary constants, $\cc{o}$ and 
$d^{\scriptstyle \sf o}$. The freedom to choose $\cc{o}$ represents a
residual gauge freedom that keeps the metric in the light-cone gauge;
this freedom is documented in Ref.~\cite{preston-poisson:06b} and
Secs.~VI A and VI C of PV. For the time being we shall keep the
residual freedom of the light-cone gauge intact, and defer the task of
making specific choices of integration constants to
Sec.~\ref{sec:horizon}. On the other hand, the freedom to choose
$d^{\scriptstyle \sf o}$ corresponds to the freedom of adding to
$\ko_1$ a solution to the homogeneous differential equation. This is
the equation that would govern an $\ell = 3$ perturbation in
first-order perturbation theory, and such a perturbation would
correspond to the presence of a tidal octupole moment $\E_{abc}$. The
freedom to adjust $d^{\scriptstyle \sf o}$ is therefore the freedom to
shift the definition of $\E_{abc}$ by a multiple of the tensor
$\K_{abc}$. This shift is meaningless, and we eliminate this freedom
by discarding the solution to the homogeneous equation. Similar
considerations apply to other radial functions, and in all cases we
eliminate the freedom to add solutions to the homogeneous equations. 

%\begingroup
%\squeezetable
\begin{table}
\caption{Radial functions appearing in the metric of
  Eq.~(\ref{metric}); $f := 1-2M/r$.} 
\begin{ruledtabular} 
\begin{tabular}{l} 
$\eq_1 = f^2$ \\ 
$\eq_4 = f$ \\ 
$\eq_7 = 1 - 2\frac{M^2}{r^2}$ \\  
$\bq_4 = f$ \\ 
$\bq_7 = 1 - 6\frac{M^2}{r^2}$ \\ 
\\ 
$\ehatq_1 = \frac{M^2}{r^2} - 4\frac{M^3}{r^3} + 2\frac{M^4}{r^4} 
+ \gam{q} \frac{M^4}{r^4}$ \\ 
$\ehatq_4 = \frac{8}{9} \frac{M^2}{r^2} - \frac{4}{9} \frac{M^4}{r^4} 
- \frac{2}{3} \gam{q} \frac{M^3}{r^3} \Bigl( 1 + \frac{M}{r} \Bigr)$
\\ 
$\ehatq_7 = \frac{7}{9} \frac{M^2}{r^2} 
- \frac{1}{3} \gam{q} \frac{M^3}{r^3}$ \\ 
$\bhatq_4 = \frac{8}{9} \frac{M^2}{r^2} 
- \frac{4}{9} \frac{M^3}{r^3}
- \frac{16}{9} \frac{M^4}{r^4}$  \\ 
$\bhatq_7 = \cc{q}\frac{M^2}{r^2} 
- \frac{2}{9} \frac{M^3}{r^3}$ \\ 
\\ 
$\kd_1 = 2\frac{M}{r} - \frac{34}{5} \frac{M^2}{r^2} 
+ \frac{32}{5} \frac{M^3}{r^3} 
+ \cc{d} \frac{M^4}{r^4}$ \\ 
$\kd_4 = \frac{M}{r} - \frac{8}{5} \frac{M^2}{r^2} 
- \frac{24}{5} \frac{M^4}{r^4} 
- \cc{d} \frac{M^4}{r^4}$ \\ 
$\fd_4 = \gam{d} \frac{M}{r} + \frac{8}{5} \frac{M^2}{r^2}$ \\ 
\\ 
$\ko_1 = 2\frac{M^2}{r^2} - \frac{8}{3} \frac{M^3}{r^3} 
+ \cc{o} \frac{M^4}{r^4}$ \\ 
$\ko_4 = \frac{4}{3} \frac{M^2}{r^2} + 4\frac{M^4}{r^4} 
+ \frac{1}{4} \cc{o} \frac{M^3}{r^3} \Bigl(5 + 2\frac{M}{r} \Bigr)$
\\  
$\ko_7 = \frac{4}{3} \frac{M^2}{r^2}  
+ \frac{1}{2} \cc{o} \frac{M^3}{r^3}$ \\
$\fo_4 = \frac{4}{3} \frac{M^2}{r^2} - \frac{10}{3} \frac{M^3}{r^3}$
\\ 
$\fo_7 = \gam{o} \frac{M^2}{r^2} + \frac{4}{3} \frac{M^3}{r^3}$ 
\end{tabular}
\end{ruledtabular} 
\label{tab:radial} 
\end{table} 
%\endgroup

With this understood, we find that the field equations give rise to
the radial functions listed in Table~\ref{tab:radial}. The residual
gauge freedom is contained in the six arbitrary constants $\gam{d}$, 
$\gam{q}$, $\gam{o}$, $\cc{d}$, $\cc{q}$, and $\cc{o}$. 

Before moving on, it is worthwhile to examine the meaning of the gauge 
constant $\gam{d}$, which will make an appearance in later sections. A
glance at Table~\ref{tab:radial} reveals that the constant is featured
in the function $\fd_4$ only, and that it gives rise to the (odd-parity) 
metric perturbation $p_{vA} = -\gam{d} M r^2 \Fd_A$, with all other 
components vanishing. If the perturbation is decomposed as in
Eq.~(\ref{odd_sector}), then $h_v = -\gam{d} M r^2 \Fd_{\sf m}$ is the
associated perturbation variable, and it is easy to show that it is
produced by a gauge transformation generated by the vector 
$\Xi_A = \sum_{\sf m} \xi_{\sf m} X^{1,{\sf m}}_A$ with
$\xi_{\sf m} = \gam{d} M r^2 \int \Fd_{\sf m}\, dv$. This represents 
a change of angular coordinates described by 
\begin{equation} 
\delta \theta^A = \gam{d} M \Omega^{AB} \int \Fd_B\, dv. 
\label{gamd_origin} 
\end{equation} 
To better understand this result, let us assume that the tidal field
is in a state of rigid rotation about the black hole's rotation axis, 
so that $\Eq_0 = \mbox{constant}$ and $\Eq_{1c} = \Eq_{1s} = 0$. These
assignments imply that $\Fd_0 = -2\chi \Eq_0 = -\chi(\E_{11} 
+ \E_{22})$, $\Fd_{1c} = \Fd_{1s} = 0$, and Eq.~(\ref{gamd_origin})
reduces to $\delta \theta = 0$ and $\delta \phi = \omega_{\rm gauge} v$,  
where $\omega_{\rm gauge} := \gam{d} \chi M (\E_{11} + \E_{22})$. We 
see that in this case, the gauge constant $\gam{d}$ is associated with 
a uniform rotation around the black hole's rotation axis. More
generally, Eq.~(\ref{gamd_origin}) describes a small precession of the
spatial coordinates around the rotation axis. 

\section{Metric in the corotating frame} 
\label{sec:corotating} 

The background metric of Eqs.~(\ref{background_metric1}) or
(\ref{background_metric2}) is presented in light-cone coordinates 
$(v,r,\theta,\phi)$ whose geometrical meaning was described in
Sec.~\ref{sec:background}. In these coordinates the event horizon is
situated at $r=2M$, and $k_\alpha := \partial_\alpha(r-2M)$ is normal
(and tangent) to it. The components $k^\alpha = (1,0,0,\omega_{\rm H})$, with 
\begin{equation} 
\omega_{\rm H} := \frac{\chi}{4M},  
\label{omega_def} 
\end{equation} 
indicate that the horizon's null generators move with an angular
velocity $d\phi/dv = \omega_{\rm H}$. We have that $\phi(v) = \phi(0) 
+ \omega_{\rm H} v$ along the generators. 

In Sec.~\ref{sec:horizon} we shall describe the intrinsic geometry of
the event horizon in the perturbed spacetime, and to prepare for this
discussion we implement a coordinate transformation that keeps the
(background) generators at a constant coordinate position. This can be
accomplished with $\theta = \vartheta$, 
$\phi = \varphi + \omega_{\rm H} v$, in which 
$\vartheta^A = (\vartheta,\varphi)$ are the new angular
coordinates. The transformation can be expressed formally as  
\begin{equation} 
\theta^A = \vartheta^A + \omega_{\rm H}^A v, \qquad 
\omega_{\rm H}^A := \omega_{\rm H} \phi^A, 
\label{transf} 
\end{equation} 
where $\phi^A = (0,1)$ is the vector introduced previously to describe
rotations around the polar axis. In Sec.~\ref{sec:hor_lock} the
transformation will be refined to ensure that the null generators of
the event horizon are at a fixed coordinate position also in the
perturbed spacetime.  

It is easy to show that to first order in $\omega_{\rm H}^A$, the metric
becomes  
\begin{equation} 
g^{\rm corot}_{vv} = g_{vv} + 2 g_{vA} \omega_{\rm H}^A, \qquad 
g^{\rm corot}_{vA} = g_{vA} + g_{AB} \omega_{\rm H}^B, \qquad 
g^{\rm corot}_{AB} = g_{AB} 
\end{equation} 
under a transformation to the corotating frame. Substitution of
Eq.~(\ref{metric}) implies that $g^{\rm corot}_{vv}$ acquires new
terms proportional to $\Eq_A \omega_{\rm H}^A$ and 
$\Bq_A \omega_{\rm H}^A$, while $g^{\rm corot}_{vA}$ acquires terms
proportional to $\Omega_{AB} \omega_{\rm H}^B$, 
$\Eq_{AB} \omega_{\rm H}^B$, and $\Bq_{AB} \omega_{\rm H}^B$. These
terms are not independent of the ones that already appear in the
metric of Eq.~(\ref{metric}). To begin, it is easy to see that the
definitions of $\omega_{\rm H}^A$ and $\chid_A$ --- refer to
Eq.~(\ref{sph_harm_decomp2}) --- imply that  
\begin{equation} 
\Omega_{AB} \omega_{\rm H}^B = -\frac{1}{4M} \chid_A. 
\end{equation} 
Similar identities are obtained by recognizing that a quantity such as
$\Eq_A \omega_{\rm H}^A = \omega_{\rm H} \Eq_\phi$ can be decomposed
in scalar harmonics $Y^{\ell\m}$, while a quantity such as 
$\Bq_{AB} \omega_{\rm H}^B  = \omega_{\rm H} \Bq_{A\phi}$ can be
decomposed in odd-parity vector harmonics $X^{\ell\m}_A$. In this way
we obtain the relations  
\begin{subequations} 
\begin{align} 
\Eq_A \omega_{\rm H}^A &= \frac{1}{8M} \chi \partial_\phi \Eq, \\ 
\Bq_A \omega_{\rm H}^A &= -\frac{3}{20M} \Kd + \frac{1}{4M} \Ko, \\ 
\Eq_{AB} \omega_{\rm H}^B &= \frac{3}{10M} \Fd_A       
+ \frac{1}{6M} \chi \partial_\phi \Eq_A 
- \frac{1}{4M} \Fo_A, \\ 
\Bq_{AB} \omega_{\rm H}^B &= -\frac{3}{10M} \Kd_A       
+ \frac{1}{6M} \chi \partial_\phi \Bq_A 
+ \frac{1}{4M} \Ko_A. 
\end{align} 
\end{subequations} 

Making the substitutions, we find that the metric in the corotating
frame takes the same form as in Eq.~(\ref{metric}), except that the
rotational term in $g^{\rm corot}_{vA}$ is now given by 
\begin{equation} 
\biggl( \frac{2M^2}{r} - \frac{r^2}{4M} \biggr) \chid_A, 
\label{rot_mod}
\end{equation} 
and that a number of radial functions are now shifted relative to
their original expression. The transformation produces    
\begin{subequations} 
\label{radial_shift} 
\begin{align} 
& \ehatq_1 \to  \ehatq_1 + \frac{1}{6} \frac{r}{M} \eq_4, \qquad 
\ehatq_4 \to  \ehatq_4 + \frac{1}{18} \frac{r}{M} \eq_7, \qquad 
\bhatq_4 \to \bhatq_4 + \frac{1}{18} \frac{r}{M} \bq_7, \\ 
& \kd_1 \to \kd_1 - \frac{1}{5} \frac{r}{M} \bq_4, \qquad 
\kd_4 \to \kd_4 - \frac{1}{10} \frac{r}{M} \bq_7, \qquad 
\fd_4 \to \fd_4 + \frac{1}{10} \frac{r}{M} \eq_7, \\ 
& \ko_1 \to \ko_1 - \frac{1}{3} \frac{r}{M} \bq_4, \qquad  
\ko_4 \to \ko_4 + \frac{1}{12} \frac{r}{M} \bq_7, \qquad
\fo_4 \to \fo_4 + \frac{1}{12} \frac{r}{M} \eq_7;
\end{align} 
\end{subequations} 
the remaining radial functions stay unchanged. Another change in
the metric is that the tidal potentials must now be expressed in terms
of $\vartheta^A$ instead of $\theta^A$; this simply involves substituting
$\vartheta$ for $\theta$, and $\varphi + \omega_{\rm H} v$ for
$\phi$. The differentiation with respect to $\phi$ becomes a
differentiation with respect to $\varphi$, but since this can still be
related to a Lie derivative in the direction of $\phi^A$ [refer to
Eq.~(\ref{Lie})], there is no pressing need to alter the notation.  
 
\section{Horizon locking} 
\label{sec:hor_lock} 

In Sec.~\ref{sec:corotating} we introduced a system of angular
coordinates $\vartheta^A$ such that in the background spacetime of
Eq.~(\ref{background_metric2}) --- with the rotational term modified to
the expression of Eq.~(\ref{rot_mod}) --- the null generators of the
event horizon are described by the parametric equations $v = v$,
$r=2M$, $\vartheta = \mbox{constant}$, and 
$\varphi = \mbox{constant}$. In this section we show that we can
choose the six undetermined constants $\gam{d}$, $\gam{q}$, $\gam{o}$,
$\cc{d}$, $\cc{q}$, and $\cc{o}$ of Sec.~\ref{sec:perturbed} in such a
way as to leave the parametric description of the horizon unchanged in
the perturbed spacetime. In other words, our selection of constants
keeps the horizon at $r = 2M$, keeps $\vartheta^A$ constant on the
null generators, and retains $v$ as a parameter on each generator.   

To show that it is indeed possible to keep the coordinate description
of the horizon unchanged in the perturbed spacetime, and to obtain the
necessary conditions to achieve this, we adapt the discussion of
Sec.~3.2 of Ref.~\cite{vega-poisson-massey:11} to a rotating black 
hole. The original discussion involved a nonrotating black hole, and
it relied on a decomposition of the perturbation in spherical
harmonics. Here we allow the background spacetime to describe a
slowly rotating black hole, and we abandon the decomposition in
spherical harmonics, which is not required for this discussion.  

We assume that the background metric $g_{\alpha\beta}^{\rm back}$ of
the slowly rotating black hole is expressed in coordinates
$(v,r,\vartheta^A)$ such that the parametric description of the
horizon is given by $v=v$, $r=2M$, $\vartheta^A = \alpha^A$, in which
$\alpha^A$ are constant generator labels. (Notice that our notation
here differs from that of Sec.~\ref{sec:perturbed}. Here the
background metric includes the rotation of the black hole. In
Sec.~\ref{sec:perturbed} the rotation was excluded from the background
metric.) In general the description of the horizon will be shifted to  
\begin{equation} 
v = v, \qquad r = 2M\bigl[ 1 + B(v,\alpha^A)\bigr], \qquad
\vartheta^A = \alpha^A + \Xi^A(v,\alpha^A)
\end{equation} 
in the perturbed spacetime with metric $g_{\alpha\beta} =  
g_{\alpha\beta}^{\rm back} + p_{\alpha\beta}$. Here $2MB$ and $\Xi^A$ 
are the components of a Lagrangian displacement vector that takes a
horizon point identified by $(v,\alpha^A)$ in the background spacetime
to a point also identified by $(v,\alpha^A)$ in the perturbed
spacetime. We keep $\alpha^A$ as constant generator labels in the 
perturbed spacetime, and we keep $v$ as parameter on the
generators. 

The vector $k^\alpha = \partial x^\alpha/\partial v$ is tangent to the
congruence of null generators, and the vectors 
$e^\alpha_A = \partial x^\alpha/\partial \alpha^A$, orthogonal to
$k^\alpha$, point from one generator to another generator. The
condition that $k^\alpha$ be null in the perturbed spacetime gives
rise to the differential equation   
\begin{equation} 
4M \frac{\partial B}{\partial v} - B + p_{vv}(v,2M,\vartheta^A) = 0
\label{B_eq} 
\end{equation} 
for the function $B(v,\alpha^A)$. On the other hand, the condition
$k_\alpha e^\alpha_A = 0$ gives rise to 
\begin{equation} 
(2M)^2 \Omega_{AB} \frac{\partial \Xi^B}{\partial v} 
+ 2 M \frac{\partial B}{\partial \alpha^A} 
+ 12 M^2 B \Omega_{AB} \omega_{\rm H}^B 
+ p_{vA}(v,2M,\vartheta^A) = 0,  
\label{Xi_eq} 
\end{equation}       
a differential equation for $\Xi^A$.  

Suppose that we wish to impose $B=0$, so that the coordinate position
of the horizon remains at $r=2M$ in the perturbed spacetime. Equation
(\ref{B_eq}) reveals that $p_{vv}$ must then vanish at
$r=2M$. Conversely, setting $p_{vv} = 0$ implies that 
$B = B(0,\alpha^A) e^{v/4M}$, which in general is incompatible with
the requirement that $B$ remain small to describe a perturbed
horizon. The only acceptable solution is $B=0$, and we conclude that a
necessary and sufficient condition for a horizon at $r=2M$ is the
first horizon-locking condition,  
\begin{equation} 
p_{vv}(v,2M,\vartheta^A) = 0.   
\label{hor_lock1} 
\end{equation} 
With $B = 0$ we next find from Eq.~(\ref{Xi_eq}) that $\Xi^A = 0$
implies that $p_{vA}$ must vanish at $r=2M$. Conversely, setting
$p_{vA} = 0$ leads to $\Xi^A = \Xi^A(\alpha^B)$, with the right-hand
side independent of $v$. Such a shift in $\alpha^A$ represents an
uninteresting constant relabeling of the null generators, and
there is no loss of generality if we simply set $\Xi^A = 0$. A
necessary and sufficient condition for the preservation of the
generator labels is therefore the second horizon-locking condition,   
\begin{equation} 
p_{vA}(v,2M,\vartheta^A) = 0. 
\label{hor_lock2} 
\end{equation} 
With these conditions we have that the parametric description of the 
horizon is given by $v=v$, $r=2M$, and $\vartheta^A  = \alpha^A$ also
in the perturbed spacetime. 

A third horizon-locking condition arises as a consequence of
Eq.~(\ref{hor_lock2}) and the vacuum Einstein field equations examined
at $r=2M$. As was shown in Secs.~VI A and VI C of Poisson and Vlasov
\cite{poisson-vlasov:10}, this condition can be formulated in terms of
the $h^{\ell\m}_v$ and $h^{\ell\m}_2$ quantities introduced in
Eq.~(\ref{odd_sector}). We have   
\begin{equation} 
h^{\ell\m}_2(v,r=2M) = -\frac{8M^2}{(\ell-1)(\ell+2)} 
\frac{\partial h^{\ell\m}_v}{\partial r} \biggr|_{r=2M}. 
\label{hor_lock3}
\end{equation} 
The third horizon-locking condition implicates only the odd-parity
sector of the perturbation. 

With the radial functions listed in Table~\ref{tab:radial} and shifted
to the corotating frame by Eq.~(\ref{radial_shift}), we find that the
horizon-locking conditions of Eqs.~(\ref{hor_lock1}),
(\ref{hor_lock2}), and (\ref{hor_lock3}) can all be enforced if we
make the assignments 
\begin{equation} 
\cc{d} = -\frac{8}{5}, \qquad 
\cc{q} = -\frac{1}{3}, \qquad 
\cc{o} = -\frac{8}{3} 
\label{const_even} 
\end{equation} 
and 
\begin{equation} 
\gam{d} = -1, \qquad 
\gam{q} = 2, \qquad 
\gam{o} = -\frac{4}{3} 
\label{const_odd}
\end{equation}  
for the undetermined constants of the perturbed metric. We recall that
the freedom to choose these six constants represents a refinement of
the light-cone gauge that preserves the geometrical meaning of the
original coordinates $(v,r,\theta,\phi)$. With the choices of
Eqs.~(\ref{const_even}) and (\ref{const_odd}), we have established
that a transformation to the corotating frame $(v,r,\vartheta,\varphi)$
produces a metric for which the parametric description of the event
horizon is also preserved: it is given by $v=v$, $r=2M$, $\vartheta =
\mbox{constant}$, and $\varphi = \mbox{constant}$ in both the
unperturbed and perturbed spacetimes.  
 
\section{Horizon geometry} 
\label{sec:horizon} 
   
As discussed at length in Ref.~\cite{vega-poisson-massey:11}, the 
geometry of the perturbed horizon can be conveniently described by
adopting $(v,\vartheta^A)$ as intrinsic coordinates. These, we recall,
are intimately related to the behavior of the horizon's generators,
and as such they provide a preferred coordinate system on the
horizon. The line element is given by
\begin{equation} 
ds^2 = \gamma_{AB}\, d\vartheta^A d\vartheta^B, 
\end{equation} 
where $\gamma_{AB} := g_{\alpha\beta} e^\alpha_A e^\alpha_B 
= g_{AB}(v,2M,\vartheta^A)$ is the horizon's intrinsic metric. A
considerable virtue of this coordinate system is that the metric is 
not merely degenerate on the null hypersurface, it is explicitly
two-dimensional. 

A simple computation involving the corotating metric of
Sec.~\ref{sec:corotating} and the assignments of
Eqs.~(\ref{const_even}) and (\ref{const_odd}) reveals that the
horizon's intrinsic metric is given by  
\begin{equation} 
\gamma_{AB} = (2M)^2 \Omega_{AB} - \frac{8}{3} M^4 \biggl[ 
\Bigl( 1 + \frac{2}{3} \chi \partial_\varphi \Bigr) 
\bigl( \Eq_{AB} + \Bq_{AB} \bigr) 
- \Fo_{AB} - \Ko_{AB} \biggr].  
\label{hor_metric} 
\end{equation}
The metric involves some of the irreducible potentials introduced in  
Sec.~\ref{sec:potentials}, now expressed in terms of $\vartheta^A$
through the combination $\vartheta^A + \omega_{\rm H}^A v$, 
which describes the transformation to the corotating frame. The metric
implicates the quadrupolar tidal potentials $\Eq_{AB}$ and $\Bq_{AB}$,
and we observe that the rotation induces an azimuthal shift that can
be expressed as 
\begin{equation} 
\Bigl( 1 + \frac{2}{3} \chi \partial_\varphi \Bigr) 
\Eq_{AB}(\vartheta,\varphi + \omega_{\rm H} v) = 
\Eq_{AB}(\vartheta,\varphi + \omega_{\rm H} v + {\textstyle \frac{2}{3}} \chi) 
+ O(\chi^2),
\label{ang_shift}
\end{equation} 
with a similar equation holding for the terms involving
$\Bq_{AB}$. The metric also implicates the octupolar potentials
$\Fo_{AB}$ and $\Ko_{AB}$ that result from the coupling of the spin
pseudovector $\chi_a$ with the tidal moments $\E_{ab}$ and $\B_{ab}$.  

The metric of Eq.~(\ref{hor_metric}) contains information about the
horizon's intrinsic geometry, and it also contains information about
the generator-anchored coordinate system that was placed on the
event horizon. Purely geometrical information can be extracted by
computing the Ricci scalar ${\cal R}$ associated with the
two-dimensional metric. A simple calculation gives
\begin{equation} 
(2M)^2 {\cal R} = 2 
- 8M^2 \Bigl( 1 + \frac{2}{3} \chi \partial_\varphi \Bigr) \Eq 
+ \frac{40}{3} M^2 \Ko, 
\end{equation} 
and we observe that the dependence on $\B_{ab}$ and $\F_{abc}$ has
disappeared --- the Ricci scalar involves irreducible potentials
of even parity only. In view of Eq.~(\ref{ang_shift}) and the definitions
introduced in Sec.~\ref{sec:potentials}, an alternative expression for
${\cal R}$ is   
\begin{equation}   
(2M)^2 {\cal R} = 2 
- 8M^2 \E_{ab} \Omega^a_\dagger \Omega^b_\dagger 
+ \frac{40}{3} M^2 \B_{\langle ab} \chi_{c\rangle} 
\Omega^a \Omega^b \Omega^c,
\label{hor_curvature}
\end{equation} 
where 
\begin{equation} 
\Omega^a := \bigl[ \sin\vartheta \cos(\varphi + \omega_{\rm H} v),
\sin\vartheta \sin(\varphi + \omega_{\rm H} v), \cos\vartheta \bigr] 
\end{equation} 
specifies the direction to a given point on the horizon, and 
\begin{equation} 
\Omega^a_\dagger := \bigl[ 
\sin\vartheta \cos(\varphi + \omega_{\rm H} v 
+ {\textstyle \frac{2}{3}} \chi), 
\sin\vartheta \sin(\varphi + \omega_{\rm H} v
+ {\textstyle \frac{2}{3}} \chi), 
\cos\vartheta \bigr] 
\end{equation} 
is its shifted version. These results were already displayed and
discussed in Sec.~\ref{sec:intro} --- refer back to 
Eq.~(\ref{hor_int_geom}) and the following discussion.  

\section{Perturbed spacetime in harmonic gauge} 
\label{sec:harmonic} 
  
In the remaining portions of this paper we shall determine the tidal 
moments $\E_{ab}$ and $\B_{ab}$ of the perturbed black-hole spacetime
by immersing the black hole in a post-Newtonian tidal environment,
thereby generalizing the work of Taylor and Poisson
\cite{taylor-poisson:08} to include spin. The first step in this
exercise is to cast the perturbed black-hole metric in a harmonic
coordinate system that will facilitate its matching to the
post-Newtonian metric; this shall be our task in this section. We
follow and generalize the method outlined in Sec.~III of Taylor and
Poisson, which consists of (i) transforming the metric perturbation
from the light-cone gauge of Sec.~\ref{sec:perturbed} to a new
harmonic gauge, keeping the background coordinates fixed to the  
$(v,r,\theta,\phi)$ system, and (ii) transforming the background
coordinates to a system of Cartesian harmonic coordinates. We adopt
the same perspective as in Sec.~\ref{sec:perturbed}, and choose the
background spacetime to have the Schwarzschild metric of
Eq.~(\ref{Schwarzschild}); the rotational and tidal terms in the
metric of the perturbed spacetime are then viewed as a first-order
perturbation, and the coupling between them as a second-order
perturbation. As before we exclude terms quadratic in $\chi$ and
quadratic in the tidal moments from the second-order perturbation.    

To define the harmonic gauge in the background coordinates
$(v,r,\theta,\phi)$, we introduce a collection of four scalar fields
$X^{(\mu)}$, 
\begin{subequations} 
\label{harmonic_fields} 
\begin{align} 
X^{(0)} &:= v - r - 2M \ln(r/2M - 1), \\ 
X^{(1)} &:= (r-M) \sin\theta\cos\phi, \\ 
X^{(2)} &:= (r-M) \sin\theta\sin\phi, \\
X^{(3)} &:= (r-M) \cos\theta,  
\end{align} 
\end{subequations} 
and demand that these be solutions to the scalar wave equation  
\begin{equation} 
g^{\alpha\beta} \nabla_\alpha \nabla_\beta X^{(\mu)} = 
\frac{1}{\sqrt{-g}} \partial_\alpha \bigl( 
g^{\alpha\beta} \partial_\beta X^{(\mu)} \bigr) = 0   
\label{harmonic_condition} 
\end{equation}  
in the perturbed spacetime with metric $g_{\alpha\beta}$ and
covariant-derivative operator $\nabla_\alpha$. A simple computation
confirms that each wave equation is satisfied in the background 
spacetime with metric $g^{\rm back}_{\rm \alpha\beta}$, and
$X^{(\mu)}$ are indeed recognized as the Cartesian harmonic
coordinates of the Schwarzschild metric. Setting $g_{\alpha\beta} 
= g^{\rm back}_{\alpha\beta} + p_{\alpha\beta}$, the inverse metric is   
$g^{\alpha\beta} = g_{\rm back}^{\alpha\beta}  
- p^{\alpha\beta} + p^{\alpha\mu} p^\beta_{\ \mu}$ to second order in 
the perturbation, and the metric determinant is $g = g_{\rm back}(1 
+ p + \frac{1}{2} p^2 - \frac{1}{2} p_{\mu\nu} p^{\mu\nu})$; indices
on $p_{\alpha\beta}$ are raised with the background metric, and $p :=
g^{\alpha\beta}_{\rm back} p_{\alpha\beta}$. Making the substitutions
in the wave equation, we find that it remains valid to second order in
the perturbed spacetime when  
\begin{equation} 
\nabla^{\rm back}_\alpha \bigl( \psi^{\alpha\beta} 
\partial_\beta X^{(\mu)} \bigr) = 0, 
\label{harm1} 
\end{equation} 
where $\nabla^{\rm back}_\alpha$ is the covariant-derivative operator
compatible with the background metric, and  
\begin{equation} 
\psi^{\alpha\beta} := p^{\alpha\beta} 
- \frac{1}{2} p\, g_{\rm back}^{\alpha\beta} 
- p^{\alpha\mu} p^\beta_{\ \mu} 
+ \frac{1}{2} p\, p^{\alpha\beta} 
- \frac{1}{8} \bigl( p^2 - 2p_{\mu\nu} p^{\mu\nu} \bigr)
g_{\rm back}^{\alpha\beta}. 
\label{harm2} 
\end{equation} 
A perturbation $p^{\rm harm}_{\alpha\beta}$ that satisfies these
conditions shall be said to be in the harmonic gauge, irrespective of
the choice of background coordinates. 

The question that faces us now is the following: Given a perturbation
$p^{\rm LC}_{\alpha\beta}$ presented in the light-cone gauge, how do
we obtain a physically equivalent perturbation 
$p^{\rm harm}_{\alpha\beta}$ that satisfies the harmonic conditions of
Eq.~(\ref{harm1})? The answer, of course, is to perform a gauge
transformation. Given a perturbation $p^{\rm old}_{\alpha\beta}$
presented in an old gauge, a second-order transformation to a new
gauge is described by (see, for example, Ref.~\cite{bruni-etal:97})    
\begin{equation} 
p^{\rm new}_{\alpha\beta} = p^{\rm old}_{\alpha\beta}  
- {\cal L}_\xi g^{\rm back}_{\alpha\beta} 
- {\cal L}_\xi p^{\rm old}_{\alpha\beta} 
+ \frac{1}{2} {\cal  L}_\xi {\cal L}_\xi g^{\rm back}_{\alpha\beta}, 
\label{gauge_transf}
\end{equation} 
in which $\xi^\alpha$ is the generating vector field, and 
${\cal L}_\xi$ indicates Lie differentiation in the direction of
$\xi^\alpha$. In our current application, $p^{\rm old}_{\alpha\beta}$
refers to the perturbation in the light-cone gauge, 
$p^{\rm new}_{\alpha\beta}$ is the perturbation in harmonic gauge, and
inserting Eq.~(\ref{gauge_transf}) within Eq.~(\ref{harm1}) gives
rise to a nonlinear, second-order differential equation for the
vector $\xi^\alpha$. Once a solution has been identified, insertion
back into Eq.~(\ref{gauge_transf}) produces the desired 
$p^{\rm harm}_{\alpha\beta}$. 

The complete gauge transformation can be built in stages, first by
finding the first-order gauge transformations that separately cast the
rotational and tidal perturbations in the harmonic gauge, and next
finding the second-order transformation that accounts for the
couplings between rotational and tidal terms. It is easy to show that
the vector $\xi_\alpha[\chi]$ that transforms the rotational
perturbation from the light-cone gauge to the harmonic gauge has 
\begin{equation} 
\xi_A[\chi] = -\frac{M^2 r}{r-M}\, \chid_A 
\label{gauge_rot} 
\end{equation} 
as nonvanishing components. The vectors $\xi_\alpha[\E]$ and 
$\xi_\alpha[\B]$ that transform the tidal perturbations from the
light-cone gauge to the harmonic gauge were identified by Taylor and
Poisson. They have 
\begin{equation} 
\xi_v[\E] = -\frac{1}{3} r^3 f \Eq, \qquad 
\xi_r[\E] = \frac{1}{3} r^3 \Eq, \qquad 
\xi_A[\E] = -\frac{r^5 f^2}{3(r-M)} \Eq_A 
\label{gauge_E} 
\end{equation} 
and 
\begin{equation} 
\xi_v[\B] = 0, \qquad \xi_r[\B] = 0, \qquad 
\xi_A[\B] = \frac{1}{3} r^2 (r^2-6M^2) \Bq_A. 
\label{gauge_B} 
\end{equation} 
The complete, second-order transformation can be expressed as 
\begin{equation} 
\xi_\alpha = \xi_\alpha[\chi] + \xi_\alpha[\E] + \xi_\alpha[\B] 
+ \xi_\alpha[\chi\E] + \xi_\alpha[\chi\B], 
\label{gauge_complete} 
\end{equation} 
where 
$\xi_\alpha[\chi \E]$ accounts for the coupling between the rotational
and even-parity tidal terms, while $\xi_\alpha[\chi \B]$ takes care of
the coupling between the rotational and odd-parity tidal terms. Each
vector can be obtained by inserting Eq.~(\ref{gauge_complete}) within
Eq.~(\ref{gauge_transf}), and that within Eq.~(\ref{harm1}), and
integrating the coupled differential equations. Because the solution
to the first-order problem has been previously identified, the
differential equations for $\xi_\alpha[\chi \E]$ and 
$\xi_\alpha[\chi \B]$ are linear, and contain source terms generated  
by the known $p^{\rm LC}_{\alpha\beta}$, $\xi_\alpha[\chi]$, 
$\xi_\alpha[\E]$, and $\xi_\alpha[\B]$. 

The rules implicated in the creation of the metric ansatz of
Eq.~(\ref{metric}) can be employed here also, and they imply that the
second-order gauge vectors can be decomposed as 
\begin{subequations} 
\label{gauge_chiE} 
\begin{align} 
\xi_v[\chi \E] &= r^3 \pq_v\, \chi \partial_\phi \Eq, \\ 
\xi_r[\chi \E] &= r^3 \pq_r\, \chi \partial_\phi \Eq, \\
\xi_A[\chi \E] &= r^4 \pq\, \chi \partial_\phi \Eq
+ r^4 \pd\, \Fd_A + r^4 \po\, \Fo_A,
\end{align} 
\end{subequations} 
and 
\begin{subequations} 
\label{gauge_chiB} 
\begin{align} 
\xi_v[\chi \B] &= r^3 \qd_v\, \Kd + r^3 \qo_v\, \Ko, \\ 
\xi_r[\chi \B] &= r^3 \qd_r\, \Kd + r^3 \qo_r\, \Ko, \\ 
\xi_A[\chi \B] &= r^4 \qd\, \Kd_A + r^4 \qo\, \Ko_A 
+ r^4 \qq\, \chi \partial_\phi \Bq_A, 
\end{align} 
\end{subequations} 
where the various coefficients $\pq_v$, $\pq_r$, $\cdots$, $\qo$,
$\qq$ are functions of $r$ that are determined by integrating the 
differential equations for $\xi_\alpha[\chi \E]$ and 
$\xi_\alpha[\chi \B]$. The solutions to these equations are linear
superpositions of particular solutions to the sourced equations and 
general solutions to the homogeneous equations. The homogeneous terms
come with a number of integration constants, and these represent the
freedom to perform an additional transformation that keeps 
$p^{\rm harm}_{\alpha\beta}$ within the harmonic gauge; we have 
used this freedom to simplify the expressions for the gauge vectors
and resulting metric perturbation to the largest extent possible. Our 
solutions are displayed in Table~\ref{tab:gauge}. We recall that the
starting point of our gauge transformation is the metric of
Eq.~(\ref{metric}), written in terms of the radial functions of
Table~\ref{tab:radial}, which involve the six arbitrary constants of
the light-cone gauge. Accordingly, the functions listed in
Table~\ref{tab:gauge} feature a dependence on these constants; this
dependence can be eliminated by imposing the choices made in
Eqs.~(\ref{const_even}) and (\ref{const_odd}), which anchor the
light-cone gauge to the null generators of the perturbed
horizon. Another point worthy of notice is that the functions $\pq_r$,
$\qd_r$, and $\qo_r$ diverge at $r=2M$: the harmonic gauge is singular
at the event horizon, in spite of the fact that the background
coordinates are well behaved there.  

%\begingroup
%\squeezetable
\begin{table}
\caption{Radial functions appearing in the gauge vectors of
  Eqs.~(\ref{gauge_chiE}) and (\ref{gauge_chiB}); $f := 1-2M/r$.}  
\begin{ruledtabular} 
\begin{tabular}{l} 
$\pd = -2 (\gam{d}+1) \frac{M^2}{r^2} \ln r 
- \frac{(r^2-6Mr+M^2)M}{r^2(r-M)} \gam{d} 
+ \frac{ (188r^3-232Mr^2+23M^2r+24M^3)M^2 }{ 30r^3(r-M)^2 } $ \\ 
$\pq_v = -\frac{(r+M)M^3}{3 r^4} \gam{q} 
- \frac{(9r^3-4Mr^2-18M^2r+16M^3)M^2}{18r^4(r-M)} $ \\  
$\pq_r = -\frac{M^3}{6 r^3} \gam{q} 
+ \frac{(6r^2+Mr-8M^2)M^2}{9r^3(r-2M)} $ \\  
$\pq = -\frac{M^3}{3r^3} \gam{q} 
+ \frac{(6r^2-11Mr+6M^2)M^3}{9r^3(r-M)^2} $ \\ 
$\po = -\frac{M^2}{2r^2} \gam{o} 
- \frac{(2r^3-2Mr^2-3M^2r+2M^3)M^2}{3r^3(r-M)^2} $ \\ 
{ } \\ 
$\qd_v = -\frac{M^4}{r^4} \cc{d} 
+ \frac{(5r-12M)M}{5r^2} $ \\ 
$\qd_r = \frac{M^4}{r^3(r-2M)} \cc{d} 
+ \frac{M^2}{5r(r-2M)} $ \\  
$\qd = \frac{M^4}{2r^3(r-M)} \cc{d}
+ \frac{(5r^2-25Mr+17M^2)M}{10r^2(r-M)} $ \\
$\qq = \frac{M^2}{r^2} \cc{q} 
- \frac{(8r^2-Mr-22M^2)M^2}{18r^3(r-M)} $\\ 
$\qo_v = \frac{(5r+2M)M^3}{12r^4} \cc{o} 
+ \frac{2M^2}{3r^2} $ \\  
$\qo_r = \frac{M^3}{12r^3} \cc{o} 
- \frac{(r-M)M^2}{3r^2(r-2M)} $ \\ 
$\qo = \frac{M^3}{4 r^3} \cc{o} $ 
\end{tabular}
\end{ruledtabular} 
\label{tab:gauge} 
\end{table} 
%\endgroup

It is now a simple matter to insert the gauge vector of
Eq.~(\ref{gauge_complete}) within Eq.~(\ref{gauge_transf}) to obtain
the metric perturbation in the harmonic gauge. The metric of the
perturbed spacetime becomes 
\begin{subequations} 
\label{metric_harm} 
\begin{align} 
g_{vv} &= -f - r^2 \eq_{vv}\, \Eq 
+ r^2 \ehatq_{vv}\, \chi \partial_\phi \Eq 
+ r^2 \kd_{vv}\, \Kd + r^2 \ko_{vv}\, \Ko, \\ 
g_{vr} &= 1 + r^2 \eq_{vr}\, \Eq 
+ r^2 \ehatq_{vr}\, \chi \partial_\phi \Eq 
+ r^2 \kd_{vr}\, \Kd + r^2 \ko_{vr}\, \Ko, \\
g_{rr} &= -2r^2 \eq_{rr}\, \Eq  
+ r^2 \ehatq_{rr}\, \chi \partial_\phi \Eq 
+ r^2 \kd_{rr}\, \Kd + r^2 \ko_{rr}\, \Ko, \\
g_{vA} &= \frac{2M^2}{r} \chid_A 
+ \frac{2}{3} r^3 \bq_{v}\, \Bq_A
+ r^3 \ehatq_v\, \chi \partial_\phi \Eq_A 
+ r^3 \fd_v\, \Fd_A + r^3 \fo_v\, \Fo_A
+ r^3 \bhatq_v\, \chi \partial_\phi \Bq_A, \\  
g_{rA} &= -\frac{(2r-M)M^2}{(r-M)^2} \chid_A 
- r^3 \eq_{r}\, \Eq_A 
- \frac{2}{3} r^3 \bq_{r}\, \Bq_A 
+ r^3 \ehatq_r\, \chi \partial_\phi \Eq_A 
+ r^3 \fd_r\, \Fd_A + r^3 \fo_r\, \Fo_A
\nonumber \\ & \quad \mbox{} 
+ r^3 \bhatq_r\, \chi \partial_\phi \Bq_A 
+ r^3 \kd_r\, \Kd_A + r^3 \ko_r\, \Ko_A, \\  
g_{AB} &= r^2\Omega_{AB} 
- r^4 \ebarq\, \Omega_{AB} \Eq - r^4 \eq\, \Eq_{AB}  
+ r^4 \ebarhatq\, \Omega_{AB} \chi \partial_\phi \Eq 
+ r^4 \ehatq\, \chi \partial_\phi \Eq_{AB} 
+ r^4 \fo\, \Fo_{AB} 
\nonumber \\ & \quad \mbox{} 
+ r^4 \bhatq\, \chi \partial_\phi \Bq_{AB} 
+ r^4 \kbard\, \Omega_{AB} \Kd 
+ r^4 \kbaro\, \Omega_{AB} \Ko 
+ r^4 \ko\, \Ko_{AB}, 
\end{align} 
\end{subequations} 
and the radial functions that appear in the metric are listed
in Table~\ref{tab:radial_harm}. We notice the large number of
functions that diverge at $r=2M$, and observe that while the gauge 
vector featured the six arbitrary constants of the light-cone gauge,
the metric depends only on the two constants $\cc{d}$ and $\gam{d}$
that appear in the dipole terms. We also note that the 
$\frac{2}{3} r^3 \bq_{v}\, \Bq_A$ term in $g_{vA}$ corrects a typo
contained in Eq.~(3.33) of Taylor and Poisson; the correct numerical
coefficient is indeed $\frac{2}{3}$ instead of $\frac{1}{3}$.  

%\begingroup
%\squeezetable
\begin{table}
\caption{Radial functions appearing in the harmonic-gauge metric of  
  Eq.~(\ref{metric_harm}); $f := 1-2M/r$.}  
\begin{ruledtabular} 
\begin{tabular}{l} 
$\eq_{vv} = f^2$ \\ 
$\eq_{vr} = f$ \\ 
$\eq_{rr} = 1$ \\ 
$\eq_r = -\frac{(3r^2-6Mr+4M^2)M}{3r(r-M)^2}$ \\ 
$\ebarq = \frac{rf^2}{r-M}$ \\ 
$\eq = \frac{(3r^2-6Mr+2M^2)M}{3r^2(r-M)}$ \\ 
$\bq_v = f$ \\ 
$\bq_r = 1$ \\ 
{ } \\ 
$\ehatq_{vv} = \frac{2(2r^2-4Mr+3M^2)M^3}{3r^4(r-M)}$ \\ 
$\ehatq_{vr} =-\frac{(r^2-3Mr+3M^2)(3r^2-4Mr+2M^2)M^2}
{3r^3(r-M)^2(r-2M)}$ \\ 
$\ehatq_{rr} =\frac{(r^6-8Mr^5+36M^2r^4-82M^3r^3+96M^4r^2
-56M^5r+12M^6)M^2}{3r^2(r-M)^4(r-2M)^2}$ \\ 
$\ehatq_v = \frac{2(7r^2-15Mr+6M^2)M^3}{9r^4(r-M)}$ \\ 
$\ehatq_r = -\frac{(15r^5-43Mr^4+15M^2r^3+48M^3r^2
-40M^4r+12M^5)M^2}{9r^3(r-M)^3(r-2M)}$ \\ 
$\ebarhatq = \frac{(5r^3-2Mr^2-10M^2r+12M^3)}{6r^3(r-M)^2}$ \\ 
$\ehatq = -\frac{5(3r^2-6Mr+4M^2)}{18r^2(r-M)^2}$ \\ 
$\bhatq_v = \frac{2(r^2+2M^2)M^3}{9r^4(r-M)}$ \\ 
$\bhatq_r = \frac{2(3r^2-Mr+M^2)M^2}{9r^3(r-M)}$ \\
$\bhatq = \frac{5M^2}{18r^2}$ \\ 
{ } \\    
$\kd_{vv} = \frac{M^4}{r^4} \cc{d} 
+ \frac{2(5r^2-12Mr+5M^2)M}{5r^3}$ \\ 
$\kd_{vr} = -\frac{M^4}{r^3(r-2M)} \cc{d} 
- \frac{2(5r^4-22Mr^3+33M^2r^2-20M^3r+5M^4)M}{5r^2(r-M)^2(r-2M)}$ \\
$\kd_{rr} = \frac{2M^4}{r^2(r-2M)^2} \cc{d} 
+ \frac{2(3r^3-12Mr^2+15M^2r-4M^3)M^2}{5r(r-M)^2(r-2M)^2}$ \\ 
$\kd_r = -\frac{M^5}{2r^2(r-M)^2(r-2M)} \cc{d} 
- \frac{(5r^3-18Mr^2+24M^2r-14M^3)M}{10r(r-M)^2(r-2M)}$ \\ 
$\kbard = \frac{M^4}{r^3(r-M)} \cc{d} 
- \frac{(5r^2-7Mr+5M^2)M}{5r^2(r-M)}$ \\ 
$\fd_v = -\frac{M}{r} \gam{d} 
- \frac{2(5r^2-8Mr+4M^2)}{5r^3(r-M)}$ \\ 
$\fd_r = \frac{(r+M)(r^2-Mr+2M^2)M}{r^2(r-M)^2} \gam{d} 
+ \frac{(10r^3-15Mr^2+23M^2r-16M^3)M^2}{5r^2(r-M)^3}$ \\ 
{ } \\
$\ko_{vv} = -\frac{2(3r^2-5Mr-M^2)M^2}{3r^4}$ \\ 
$\ko_{vr} =\frac{2(3r^4-12Mr^3+14M^2r^2-3M^3r
-M^4)M^2}{3r^3(r-M)^2(r-2M)}$ \\ 
$\ko_{rr} =-\frac{2(3r^4-12Mr^3+13M^2r^2
-2M^4)M^2}{3r^2(r-M)^2(r-2M)^2}$ \\ 
$\ko_r = \frac{(r-M)M^2}{r^2(r-2M)}$ \\ 
$\kbaro = -\frac{2(r+M)M^2}{3r^3}$ \\ 
$\ko = \frac{4M^2}{3r^2}$ \\ 
$\fo_v = \frac{2(r-3M)M^3}{3r^3(r-M)}$ \\ 
$\fo_r = -\frac{(5r^2-11Mr+2M^2)M^3}{3r^2(r-M)^3}$ \\ 
$\fo = -\frac{2M^4}{3r^2(r-M)^2}$ 
\end{tabular}
\end{ruledtabular} 
\label{tab:radial_harm} 
\end{table} 
%\endgroup

Now that the perturbed metric has been recast in a harmonic gauge in
the background coordinates $(v,r,\theta,\phi)$, the remaining task is
to perform a transformation from these coordinates to Cartesian 
harmonic coordinates $t = X^{(0)}, x = X^{(1)}, y = X^{(2)}, 
z= X^{(3)}$. We carry this out in two stages. First, we transform the
metric from the $(v,r,\theta,\phi)$ coordinates to  
$(t,\bar{r},\theta,\phi)$ coordinates, with $v = t + r 
+ 2M\ln(r/2M-1)$ and $r = \bar{r}+ M$. This produces
\begin{equation} 
g_{tt} = g_{vv}, \qquad 
g_{t\bar{r}} = \frac{1}{f} g_{vv} + g_{vr}, \qquad 
g_{\bar{r}\bar{r}} = \frac{1}{f^2} g_{vv} + \frac{2}{f} g_{vr} 
+ g_{rr} 
\end{equation} 
and 
\begin{equation} 
g_{tA} = g_{vA}, \qquad 
g_{\bar{r}A} = \frac{1}{f} g_{vA} + g_{rA}, \qquad 
g_{AB} = g_{AB}.  
\end{equation} 
In the second stage we introduce Cartesian coordinates $x^a$ that are
related in the usual way to the spherical polar coordinates 
$(\bar{r},\theta,\phi)$. We express the relationship as $x^a =
\bar{r}\Omega^a(\theta^A)$, in which 
$\Omega^a := [\sin\theta\cos\phi, \sin\theta\sin\phi, \cos\theta]$. 
The transformation gives 
\begin{equation} 
g_{tt} = g_{tt}, \qquad 
g_{ta} = g_{t\bar{r}} \Omega_a 
+ \frac{1}{\bar{r}} g_{tA} \Omega^A_a, \qquad 
g_{ab} = g_{\bar{r}\bar{r}} \Omega_a \Omega_b 
+ \frac{1}{\bar{r}} g_{\bar{r}A} \bigl( \Omega_{a} \Omega^A_{b}
+ \Omega^A_a \Omega_b \bigr) 
+ \frac{1}{\bar{r}^2} g_{AB} \Omega^A_a \Omega^B_b,  
\label{metric_harm_cart} 
\end{equation} 
where $\Omega^A_a := \delta_{ab} \Omega^{AB} \partial_B \Omega^b
= \bar{r}\partial \theta^A/\partial x^a$. The
manipulations that lead to Eq.~(\ref{metric_harm_cart}) from 
Eq.(\ref{metric_harm})  are straightforward, but they produce very
long expressions for the metric components. 

Fortunately, only the components $g_{tt}$ and $g_{ta}$ are required in 
the matching of the black-hole metric with the post-Newtonian metric
to be introduced in the following section. And because the matching
will be carried out at the first-and-a-half post-Newtonian ($1.5\PN$)
approximation, the expressions for $g_{tt}$ and $g_{ta}$ can be
simplified by discarding all terms that occur at higher post-Newtonian
orders. The rules of truncation are simple, and to formulate them we
recall from Sec.~\ref{sec:intro} that $\E_{ab}$ scales as $m_2/b^3$,
$\B_{ab}$ scales as $m_2 u/b^3$, and $\chi \ll 1$ is a scale-free
quantity; here $m_2$ is a mass scale for the external matter, $b$ is a
length scale for its distance to the black hole, and $u$ is a velocity
scale. These scalings imply that $\E_{ab}$ is of Newtonian order
($0\PN$), while $\B_{ab}$ is of $0.5\PN$ order because of the 
additional factor of $u$; the tidal moments, however, also contain
higher-order post-Newtonian corrections.    

Let us examine $g_{tt}$. A term proportional to $M/\bar{r}$ in
$g_{tt}$ is a Newtonian term, and each additional factor of
$M/\bar{r}$ increases the post-Newtonian order by one unit. The
leading tidal term proportional to $\bar{r}^2 \E_{ab}$ is a Newtonian
term, and again, each additional factor of $M/\bar{r}$ gives rise to a
higher post-Newtonian correction. Turning next to the couplings
between rotational and tidal terms, we see that 
$\bar{r}^2 \chi \partial_\phi \Eq$ would be of Newtonian order, but
that the radial function $\ehatq_{vv}$, which leads off at order
$(M/\bar{r})^3$, promotes it to a $3\PN$ term that can be
neglected. Similarly, $\bar{r}^2 \Kd$ and $r^2 \Ko$ lead off at
$0.5\PN$ order, and here the radial functions promote the dipole term
to $1.5\PN$ order (this we keep), while the octupole term is promoted
to $2.5\PN$ order (and can be neglected). 

For $g_{ta}$, tradition dictates that the post-Newtonian counter must
be increased by a half unit. The leading term is proportional to  
$\bar{r}^2 \B_{ab}$, which is declared to be of $1\PN$ order; each
additional factor of $M/\bar{r}$ increases the post-Newtonian order by
one unit. The rotational term occurs at $1.5\PN$ order, and the
post-Newtonian order of the coupling terms can be determined by
adapting the rules spelled out previously for $g_{tt}$; the conclusion
is that the only contribution at $1.5\PN$ order comes from the dipole
term proportional to $\bar{r}^2 \Fd_a$.  

From these considerations we arrive at 
\begin{subequations} 
\label{metric_harm_PN1} 
\begin{align} 
g_{tt} &= -1 + \frac{2M}{\bar{r}} - \frac{2M^2}{\bar{r}^2} 
- \bar{r}^2 \biggl( 1 - \frac{2M}{\bar{r}} \biggr) \Eq 
+ 2M \bar{r}\, \Kd, \\ 
g_{ta} &= \frac{2M^2}{\bar{r}^2} \chid_a 
+ \frac{2}{3} \bar{r}^2\, \Bq_a 
- \gam{d} M \bar{r}\, \Fd_a 
\end{align} 
\end{subequations} 
for the relevant components of the metric truncated through $1.5\PN$  
order. Inserting the definitions for the tidal potentials, restoring
the factors of $G$ and $c$ that were previously set equal to unity,
and replacing the dimensionless spin $\chi^a$ with $S^a = (GM^2/c)
\chi^a$, we obtain the explicit expression 
\begin{subequations} 
\label{metric_harm_PN2} 
\begin{align} 
g_{tt} &= -1 + \frac{2GM}{c^2 \bar{r}}
-2 \biggl(\frac{GM}{c^2 \bar{r}} \biggr)^2 
- \frac{1}{c^2} \biggl( 1 - \frac{2GM}{c^2 \bar{r}} \biggr) 
\E_{ab} x^a x^b 
+ \frac{2}{c^4} \B_{ap} \hat{S}^p x^a , \\ 
g_{ta} &= \frac{2G}{c^3} \frac{(\bm{x} \times \bm{S})_a}{\bar{r}^3} 
+ \frac{2}{3c^3} \epsilon_{abp} \B^p_{\ c} x^b x^c  
- \frac{\gam{d}}{c^3} \epsilon_{abp} \E^p_{\ q} \hat{S}^q x^b 
\end{align} 
\end{subequations} 
for the black-hole metric, where $\hat{S}^a := S^a/M$ and 
$(\bm{x} \times \bm{S})_a := \epsilon_{abc} x^b S^c$. This is our
final result in this section: We have the metric of a slowly rotating
black hole perturbed by a tidal environment, expressed in Cartesian
harmonic coordinates, and truncated through $1.5\PN$ order. It should
be noted that the expression of the metric in terms of $S^a$ instead
of $\chi^a$ eliminates some factors of $c^{-1}$ that were present
implicitly in Eq.~(\ref{metric_harm_PN1}). As a result, the
post-Newtonian order of each term involving $S^a$ has been altered by
a half unit, and we now find that $1.5\PN$ terms no longer appear
explicitly in the metric of Eq.~(\ref{metric_harm_PN2}). It is
interesting to note that at this post-Newtonian order, the metric
features a dependence on the light-cone gauge constant $\gam{d}$; the
dipole term in $g_{ta}$ therefore reflects a choice of gauge, and we
recall that the meaning of this gauge freedom was explained near the
end of Sec.~\ref{sec:perturbed} --- refer back to the discussion
surrounding Eq.~(\ref{gamd_origin}).     

\section{Tidal moments for a slowly rotating black hole} 
\label{sec:tidal} 

In this section the black hole is imagined to be a member of a
post-Newtonian system that could contain one or more companion bodies,
or a continuous distribution of matter that is taken to be well
separated from the black hole; the immediate vicinity of the black
hole is assumed to be empty of matter. The tidal moments $\E_{ab}$ and 
$\B_{ab}$ shall be determined by matching the black-hole metric of
Eq.~(\ref{metric_harm_PN2}) to the post-Newtonian metric that
describes the weak mutual gravity between the black hole and the
external matter. The black-hole metric is valid in a neighborhood of
the black hole, where $\bar{r} < r_{\rm max}$ with 
$r_{\rm max} \ll b$, and the post-Newtonian metric is valid in
a region that excludes the black hole, where $\bar{r} > r_{\rm min}$
with $r_{\rm min} \gg GM/c^2$. When $GM/c^2 \ll b$ it is possible to
identify an overlap region described by $r_{\rm min} < \bar{r} <
r_{\rm max}$, in which both descriptions are valid (refer back to
Fig.~\ref{fig:f2}); the matching of the black-hole and post-Newtonian
metrics can be carried out in this region, and among other important
pieces of information, the procedure returns expressions for the tidal
moments.      

The matching procedure is described in great detail in Taylor and
Poisson \cite{taylor-poisson:08}, and their work can be directly
imported here with very few alterations to account for the black-hole
spin. We provide here a concise discussion that highlights the main
differences with the previous work.  

The starting point (Sec.~IV of Taylor and Poisson) is the
post-Newtonian metric 
\begin{subequations} 
\label{metric_PN} 
\begin{align} 
g_{tt} &= -1 + \frac{2}{c^2} U + \frac{2}{c^4} (\Psi - U^2) 
+ O(c^{-5}), \\ 
g_{ta} &= -\frac{4}{c^3} U_a + O(c^{-5})    
\end{align} 
\end{subequations} 
that describes the mutual gravity between the black hole and the
external matter. The metric is presented in a harmonic coordinate
system $(t,x^a)$ attached to the system's barycenter, and it involves
a Newtonian potential $U$, a vector potential $U_a$, and a
post-Newtonian potential $\Psi = \psi + \frac{1}{2} \partial_{tt} X$
conveniently decomposed in terms of another potential $\psi$ and a
superpotential $X$. Because the region of space surrounding the 
black hole is empty of matter, the potentials there satisfy the vacuum
field equations $\nabla^2 U = 0$, $\nabla^2 U_a = 0$, 
$\nabla^2 \psi = 0$, and $\nabla^2 X = 2U$, where $\nabla^2$ is the
familiar Laplacian operator of three-dimensional flat space; the
harmonic coordinate condition implies that the Newtonian and vector
potentials are tied by the gauge condition 
$\partial_t U + \partial_a U^a = 0$. 

The potentials must reflect the presence of a black hole near the
world line described by $\bm{z}(t)$, and they must reflect the
presence of the external matter. The matching procedure compels us to
model the black hole as an object possessing both mass and spin, but
no higher-order multipole moments. The potentials are given by 
\begin{subequations} 
\label{pNpotentials1} 
\begin{align} 
U(t,\bm{x}) &= \frac{GM}{R} + U_{\rm ext}(t,\bm{x}), \\ 
U^a(t,\bm{x}) &= -\frac{G(\bm{R} \times \bm{S})^a}{2R^3} 
+ \frac{GMu^a}{R} + U^a_{\rm ext}(t,\bm{x}), \\     
\psi(t,\bm{x}) &= \frac{GM\mu}{R} + \frac{G \bm{p} \cdot \bm{R}}{R^3} 
+ \psi_{\rm ext}(t,\bm{x}), \\ 
X(t,\bm{x}) &= GM R + X_{\rm ext}(t,\bm{x}), 
\end{align} 
\end{subequations} 
where the black-hole terms are clearly distinguished from the external
terms. With this we find that 
\begin{equation} 
\Psi(t,\bm{x}) = \frac{GM}{R} \biggl( \mu + \frac{1}{2} u^2 \biggr) 
- \frac{GM}{2R^3} (\bm{u} \cdot \bm{R})^2 
- \frac{GM}{2R} \bm{a} \cdot \bm{R} 
+ \frac{G\bm{p} \cdot \bm{R}}{R^3} 
+ \Psi_{\rm ext}(t,\bm{x}), 
\label{pNpotentials2} 
\end{equation} 
where $\Psi_{\rm ext} := \psi_{\rm ext} 
+ \frac{1}{2} \partial_{tt} X_{\rm ext}$.  We have introduced 
$\bm{R} := \bm{x}-\bm{z}(t)$, $R := |\bm{R}|$, 
$\bm{u}(t) := d\bm{z}/dt$, $\bm{a}(t) := d\bm{u}/dt$, and 
$u^2 := \bm{u} \cdot \bm{u}$. The mass monopole moment $M$ is
identified with the black-hole mass, and the current dipole moment
$\bm{S}$ is identified with its spin; $\mu(t)$ is a post-Newtonian
correction to the monopole moment, and $\bm{p}(t)$ is a post-Newtonian 
dipole moment that must be included to ensure a successful matching to
the black-hole metric. It is easy to verify that the potentials of
Eq.~(\ref{pNpotentials1}) satisfy the vacuum field equations and the
harmonic gauge condition (which requires the presence of $u^a$ in the
vector potential).  

The post-Newtonian multipoles $\mu(t)$ and $\bm{p}(t)$ are not
determined by the field equations, because the validity of the
post-Newtonian metric does not extend beyond $R = r_{\rm min}$. If the
black hole were replaced by a material body with weak internal
gravity, integrating the field equations with matter (refer to
Secs.~9.2.2 and 9.5.8 of Poisson and Will \cite{poisson-will:14})
would return the same potentials with 
$\mu = \frac{3}{2} u^2 - U_{\rm ext}(t,\bm{z})$ and 
$\bm{p} = \frac{1}{2}(3+\lambda) \bm{u} \times \bm{S}$, where
$\lambda$ is a dimensionless quantity that parametrizes the choice of
representative world line to describe the body's center-of-mass; the
choice $\lambda = 1$ can be described as ``imposing the covariant
spin supplementary condition.'' In the sequel we shall verify that
matching the black-hole and post-Newtonian metrics produces the same
expressions for $\mu$ and $\bm{p}$; we shall also see that the choice
$\lambda=1$ is natural and gives rise to important simplifications.    

To carry out the matching it is necessary to transform the
post-Newtonian metric from the barycentric frame $(t,x^a)$ to
another harmonic frame $(\bar{t},\bar{x}^a)$ that is at all times
centered on the black hole; we shall refer to this new frame as the
black-hole frame. The transformation was first described in the
context of weakly gravitating bodies by Kopeikin \cite{kopeikin:88},
Brumberg and Kopeikin \cite{brumberg-kopeikin:89}, and Damour, Soffel,
and Xu \cite{damour-soffel-xu:91}, then extended to compact bodies by
Racine and Flanagan \cite{racine-flanagan:05}; its details are
reviewed in Sec.~V of Taylor and Poisson. The end result is the 
following expressions for the transformed potentials,    
\begin{subequations} 
\label{pNpotentials_barred} 
\begin{align} 
\bar{U} &= \frac{GM}{\bar{r}} + \mbox{}_0 U 
+ \mbox{}_1 U_a\, \bar{x}^a + \mbox{}_2 U_{ab}\, \bar{x}^a \bar{x}^b 
+ O(\bar{r}^3), \\ 
\bar{U}^a &= -\frac{G(\bm{\bar{x}} \times \bm{S})^a}{2 \bar{r}^3} 
+ \mbox{}_0 U^a + \mbox{}_1 U^a_{\ b}\, \bar{x}^b 
+ \mbox{}_2 U^a_{\ bc}\, \bar{x}^b \bar{x}^c 
+ O(\bar{r}^3), \\ 
\bar{\Psi} &= \frac{G \bm{q} \cdot \bm{\bar{x}}}{\bar{r}^3} 
+ \frac{GM}{\bar{r}} (\mu+\dot{A}-2v^2) + \mbox{}_0 \Psi  
+ \mbox{}_1 \Psi_a\, \bar{x}^a + \mbox{}_2 \Psi_{ab}\, \bar{x}^a \bar{x}^b  
+ O(\bar{r}^3),
\end{align} 
\end{subequations} 
where $\bar{r}^2 := \delta_{ab} \bar{x}^a \bar{x}^b$, 
$\bm{q} := \bm{p} - 2 \bm{u} \times \bm{S}  - M (\bm{H} - A \bm{u})$,
and all other symbols are defined in Taylor and Poisson. It is
important to note that spin terms appear in the singular pieces of the 
potentials; for example, the spin vector $\bm{S}$
now appears within $\bar{U}^a$ [compare with Eq.~(5.34) of Taylor
and Poisson] and it appears also, along with the dipole moment
$\bm{p}$, in the expression for $\bm{q}$ [compare with Eq.~(5.35) of
Taylor and Poisson]. It is equally important to note that the spin
terms do not appear in the regular pieces, which are listed
explicitly in Eqs.~(5.36)--(5.44) of Taylor and Poisson.  

The transformed potentials are next inserted into the transformed
metric, which is now ready to be compared with the black-hole metric
of Eq.~(\ref{metric_harm_PN2}). The metrics must agree, and the
comparison gives rises to matching conditions that determine the 
unknown functions that enter the transformation between the
$(t,x^a)$ and $(\bar{t},\bar{x}^a)$ coordinates, the unknown
functions $\mu$ and $\bm{p}$ that enter the post-Newtonian metric, and
finally, the tidal moments $\bar{\E}_{ab}$ and $\bar{\B}_{ab}$. (We
place an overbar on the tidal moments to indicate that these are
defined in the black-hole frame; this notation differs from the one
adopted in earlier sections of this paper. Below we shall also
introduce tidal moments $\E_{ab}$ and $\B_{ab}$ that are defined  
in the barycentric frame of the post-Newtonian spacetime.) 
The details of the matching are described in Sec.~VI of Taylor and
Poisson, and very few alterations are required to account for the
black-hole spin.  

The matching conditions that do acquire spin terms are  
\begin{subequations} 
\label{matching_conditions} 
\begin{align} 
\mbox{}_1 U_a + \frac{1}{c^2} \bigl( \mbox{}_1 \Psi_a 
- \bar{\B}_{ab} \hat{S}^b \bigr) &= O(c^{-4}), \\ 
\bm{p} - 2\bm{u} \times \bm{S} - M(\bm{H} - A \bm{u}) &= O(c^{-2}), 
\\
\mbox{}_1 U_{ab} - \frac{1}{4} \gam{d} \epsilon_{abp} \bar{\E}^p_{\ q}
\hat{S}^q &= O(c^{-2}), 
\end{align} 
\end{subequations} 
which replace Eqs.~(6.7), (6.8), and (6.13) of Taylor and Poisson,
respectively. The second equation is the only matching condition that
permits the determination of $\bm{H}$, which appears in the coordinate
transformation, and $\bm{p}$, which appears in the post-Newtonian
metric. In the absence of spin $\bm{p}$ would vanish identically, and
the matching condition would produce $\bm{H} = A \bm{u} + O(c^{-2})$. 
In the presence of spin it is natural to keep this relation unaltered,
and to separately impose  
\begin{equation} 
\bm{p} = 2 \bm{u} \times \bm{S}. 
\label{p_determined} 
\end{equation} 
While this choice is natural, other possibilities cannot be
excluded. For example, $\bm{p}$ could be generalized to
$\frac{1}{2}(3+\lambda) \bm{u} \times \bm{S}$ at the cost of altering
the expression for $\bm{H}$; as was mentioned previously, this 
would generate the freedom to shift the representative world line
taken to represent the black hole's center-of-mass in the
post-Newtonian metric. For convenience and simplicity we prefer to
leave $\bm{H}$ unaffected by spin terms, and to stick with $\bm{p}$ as
determined by Eq.~(\ref{p_determined}). 

The spin-induced changes in the matching conditions impact only some
of the results obtained by Taylor and Poisson. Among the results that
are not changed are the expression for $\mu$ given by their Eq.~(6.24)
--- the expected $\mu = \frac{3}{2} v^2 - U_{\rm ext}$ --- and the
expressions for the tidal moments given by their Eqs.~(6.36)--(6.37);
we have    
\begin{align} 
\E_{ab} &= -\partial_{ab} U_{\rm ext} + \frac{1}{c^2} 
\biggl[ -\partial_{\langle ab \rangle} \Psi_{\rm ext} 
+ 4u^c \bigl( \partial_{ab} U^{\rm ext}_c 
- \partial_{c\langle a} U^{\rm ext}_{b\rangle} \bigr) 
- 4 \partial_{t\langle a} U^{\rm ext}_{b\rangle} 
- 2(u^2 - U_{\rm ext}) \partial_{ab} U_{\rm ext} 
\nonumber \\ & \quad \mbox{} 
+ 3 u^c u_{\langle a} \partial_{b\rangle c} U_{\rm ext} 
+ 2 u_{\langle a} \partial_{b\rangle t} U_{\rm ext} 
+ 3 \partial_{\langle a} U_{\rm ext} \partial_{b\rangle} U_{\rm ext} 
\biggr] + O(c^{-4}) 
\label{EpN} 
\end{align} 
and 
\begin{equation} 
\B_{ab} = 2 \epsilon_{pq(a} \partial^p_{\ b)} \bigl( 
U^q_{\rm ext} - u^q U_{\rm ext} \bigr) + O(c^{-2}), 
\label{BpN}
\end{equation} 
where the potentials are evaluated at $\bm{x}=\bm{z}(t)$ after
differentiation, and indices within angular brackets are symmetrized
and made tracefree; for example $u_{\langle a} \partial_{b\rangle} =
\frac{1}{2} (u_a \partial_b + u_b \partial_a) - \frac{1}{3}
\delta_{ab} u^c \partial_c$. 

Among the results that do change to account for spin are the black
hole's equations of motion, which become  
\begin{align} 
a_a &= \partial_a U_{\rm ext} + \frac{1}{c^2} \biggl[ 
\partial_a \Psi_{\rm ext} - 4 \bigl( \partial_a U^{\rm ext}_b 
- \partial_b U^{\rm ext}_a \bigr) u^b 
+ 4 \partial_t U^{\rm ext}_a 
+ (u^2 - 4 U_{\rm ext}) \partial_a U_{\rm ext} 
\nonumber \\ & \quad \mbox{} 
- u_a \bigl( 4u^b \partial_b U_{\rm ext} 
+ 3 \partial_t U_{\rm ext} \bigr) 
- \B_{ab} \hat{S}^b \biggr] + O(c^{-4}) 
\label{acceleration} 
\end{align} 
and which differ from Eq.~(6.30) of Taylor and Poisson by the presence
of the Mathisson-Papapetrou spin force \cite{mathisson:10,
  papapetrou:51a, papapetrou:51b} proportional to $\B_{ab} \hat{S}^b$.  

The tidal moments of Eqs.~(\ref{EpN}) and (\ref{BpN}) are defined in
the barycentric frame $(t,x^a)$ of the post-Newtonian metric. The
transformation to the black-hole frame $(\bar{t},\bar{x}^a)$ is
described by Eqs.~(6.33)--(6.35) of Taylor and Poisson. We have 
\begin{equation} 
\bar{\E}_{ab}(\bar{t}) = 
{\cal N}_a^{\ c}(t) {\cal N}_b^{\ d}(t) \E_{cd}(t), \qquad 
\bar{\B}_{ab}(\bar{t}) =
{\cal N}_a^{\ c}(t) {\cal N}_b^{\ d}(t) {\B}_{cd}(t),  
\end{equation} 
where 
\begin{equation} 
{\cal N}_{ab}(t) := \delta_{ab} 
- \frac{1}{c^2} \epsilon_{abc} R^c(t) + O(c^{-4}) 
\label{N_def} 
\end{equation} 
describes a post-Newtonian precession of the black-hole frame relative
to the barycentric frame. The relation between the time coordinates is
given by $\bar{t} = t - c^{-2} A(t) + O(c^{-4})$, with $A(t)$ determined by 
\begin{equation} 
\frac{dA}{dt} = \frac{1}{2} u^2 + U_{\rm ext}(t,\bm{z});   
\label{Adot} 
\end{equation} 
this equation is unchanged with respect to Eq.~(6.35) of Taylor and 
Poisson. The precession vector $\bm{R}(t)$ is determined by
\begin{equation} 
\epsilon_{abc} \frac{dR^c}{dt} = -4 \partial_{[a} 
  U^{\rm ext}_{b]} -3 u_{[a} \partial_{b]} U_{\rm ext}
-\gam{d} \epsilon_{abc} \E^c_{\ p} \hat{S}^p. 
\label{Rdot} 
\end{equation} 
Notice that this equation acquires a spin term that was not present in
Eq.~(6.35) of Taylor and Poisson. We observe that the precession of
the black-hole frame features a dependence on the gauge constant
$\gam{d}$, which was already associated with precession effects in the
discussion surrounding Eq.~(\ref{gamd_origin}).  

The conclusion of this section is that while the tidal moments of
Eqs.~(\ref{EpN}) and (\ref{BpN}) do not depend explicitly on the
black-hole spin $\bm{S}$, the acceleration vector $\bm{a}$ and the  
precession vector $\bm{R}$ both do. This implies that
$\bar{\E}_{ab}$ and $\bar{\B}_{ab}$ possess an implicit dependence
upon $\bm{S}$ provided by the motion and precession of the black-hole 
frame.  

\section{Tidal moments for a two-body system} 
\label{sec:2body} 

In this section we specialize the results of the preceding section to
a post-Newtonian system consisting of two bodies, the black hole and a
companion. We adapt the notation to this specific situation: the black
hole's mass, spin, position, velocity, and acceleration will now be
denoted $m_1$, $\bm{S}_1$, $\bm{z}_1$, $\bm{u}_1$, and $\bm{a}_1$,
respectively, and the quantities associated with the companion will be
$m_2$, $\bm{S}_2$, $\bm{z}_2$, $\bm{u}_2$, and $\bm{a}_2$. We recall
that $\hat{\bm{S}}_1 := \bm{S}_1/m_1$, 
$\hat{\bm{S}}_2 := \bm{S}_2/m_2$, and we also introduce 
$\bm{b} := \bm{z}_1 - \bm{z}_2$ and $\bm{u} := \bm{u}_1
- \bm{u}_2$. Finally, we let $b := |\bm{b}|$ be the distance
between the black hole and companion, and $\bm{n} := \bm{b}/b$ be a 
unit vector pointing from the companion to the black hole. The
positions, velocities, and accelerations all refer to the barycentric
frame $(t,x^a)$ of the post-Newtonian metric.  

The black hole's acceleration $\bm{a}_1$, given by
Eq.~(\ref{acceleration}), and the barycentric tidal moments $\E_{ab}$
and $\B_{ab}$, given by Eqs.~(\ref{EpN}) and (\ref{BpN}), can be
calculated from the external potentials $U_{\rm ext}$, 
$U^a_{\rm ext}$, and $\Psi_{\rm ext}$, which are now taken to be those
generated by the companion. Whether this is a black hole, a material 
compact body, or a weakly self-gravitating body, the external
potentials will take the same form as the black-hole terms in
Eqs.~(\ref{pNpotentials1}) and (\ref{pNpotentials2}). We therefore have  
\begin{subequations} 
\label{extpotentials} 
\begin{align} 
U_{\rm ext}(t,\bm{x}) &= \frac{Gm_2}{R_2}, \\ 
U_{\rm ext}^a(t,\bm{x}) &= 
-\frac{G(\bm{R}_2 \times \bm{S}_2)^a}{2R_2^3}  
+ \frac{Gm_2 u_2^a}{R_2}, \\     
\Psi_{\rm ext}(t,\bm{x}) &= 
\frac{Gm_2}{R_2} \biggl(  2u_2^2 - \frac{Gm_1}{b} \biggr) 
- \frac{Gm_2}{2R_2^3} (\bm{u}_2 \cdot \bm{R}_2)^2 
- \frac{G m_2}{2R_2} \bm{a}_2 \cdot \bm{R}_2  
+ \frac{2G (\bm{u}_2 \times \bm{S}_2) \cdot \bm{R}_2}{R_2^3}, 
\end{align} 
\end{subequations} 
where $\bm{R}_2 := \bm{x} - \bm{z}_2$ and $R_2 := |\bm{R}_2|$. In 
$\Psi_{\rm ext}$ we have inserted the appropriate expressions for
$\mu$ and $\bm{p}$, and in future manipulations we shall 
also substitute the appropriate expression $\bm{a}_2 = Gm_1
\bm{n}/b^2 + O(c^{-2})$ for the acceleration of the companion body. 

We first involve the external potentials in a calculation of the
barycentric tidal moments $\E_{ab}$ and $\B_{ab}$. The contributions
that are independent of spin were previously computed by Taylor and
Poisson \cite{taylor-poisson:08} --- see their Eqs.~(7.10) and (7.11)
--- and we shall focus here on the contributions arising from the spin
terms in the external potentials. Going through these calculations it
is convenient to go back and forth between the spin vector and tensor
of each body, using the relations  
\begin{equation} 
S^{ab} = \epsilon^{abc} S_{c}, \qquad 
S^a = \frac{1}{2} \epsilon^{abc} S_{bc}. 
\label{spin-vector-tensor} 
\end{equation} 
The tensorial notation, in particular, is helpful to simplify the spin
part of the external potentials, which can be expressed as 
\begin{equation} 
U^a_{\rm ext}[\text{spin}] = \frac{1}{2} G S_2^{ab} 
\partial_b \frac{1}{R_2}, \qquad
\Psi_{\rm ext}[\text{spin}] = -2 G S_2^{ab} u_{2\, b}\, 
\partial_a \frac{1}{R_2}.   
\end{equation} 
This form allows the various derivatives of the potentials to be
related to derivatives of $R^{-1}_2$, and therefore to be expressed in
terms of symmetric tracefree tensors after evaluation at
$\bm{x}=\bm{z}_1$. For example, $\partial_{ab} R^{-1}_2$ becomes 
$3n_{\langle ab \rangle}/b^3$, and $\partial_{abc} R^{-1}_2$ becomes   
$-15 n_{\langle abc \rangle}/b^4$. After straightforward manipulations
we find that the spin contributions to the barycentric tidal moments
are given by   
\begin{equation} 
\E_{ab}[\text{spin}] = \frac{15 G}{c^2 b^4} \Bigl( 
S_{2\, a}^{\ \ \, c} n_{\langle bcd \rangle} 
+ S_{2\, b}^{\ \ \, c} n_{\langle acd \rangle} 
- 2 S_{2\, d}^{\ \ \, c} n_{\langle abc \rangle} \Bigr) u^d 
\label{Espin} 
\end{equation} 
and 
\begin{equation} 
\B_{ab}[\text{spin}] = -\frac{15 G}{b^4} S_2^c n_{\langle abc \rangle}. 
\label{Bspin} 
\end{equation} 
These contributions to the tidal moments can be added to those listed
in Eqs.~(7.10) and (7.11) of Taylor and Poisson.     

We next involve the external potentials in a calculation of the black
hole's acceleration $\bm{a}_1$. We focus our attention on the spin
terms coming both from the external potentials (terms proportional to
$\bm{S}_2$) and the Mathisson-Papapetrou spin force (terms
proportional to $\bm{S}_1$ and to the product of each spin), and show 
that they reproduce the well-known expressions for the leading-order
spin-orbit and spin-spin forces. It is known that the other terms in
$\bm{a}_1$ reproduce the standard Newtonian and post-Newtonian terms
--- see Eq.~(9.245) of Poisson and Will \cite{poisson-will:14} --- in
the equations of motion.       

The spin terms arising from the external potentials in
Eq.~(\ref{acceleration}) are calculated using the methods outlined
previously for the computation of the tidal moments. The spin terms 
arising from the Mathisson-Papapetrou spin force are obtained from  
\begin{equation} 
\B_{ab} \hat{S}^b_1 = \bigl( 2 \hat{S}_1^{bc} \partial_{ab} 
+ \hat{S}_{1\, a}^{\ \ \, b} \partial_{bc} \bigr) 
\bigl( U^c_{\rm ext} - u_1^c U_{\rm ext} \bigr), 
\end{equation} 
which follows from Eq.~(\ref{BpN}) after involving
Eq.~(\ref{spin-vector-tensor}) and rearranging the products of
permutation symbols. Substituting the appropriate expressions for the
external potentials gives 
\begin{equation} 
\B_{ab} \hat{S}^b_1 = -\frac{3Gm_2}{b^3} \bigl( 
\hat{S}_{1\, a}^{\ \ \, b} n_{\langle bc \rangle} 
+ 2 \hat{S}_{1}^{bc} n_{\langle ab \rangle} \bigr) u^c 
- \frac{15 G m_2}{b^4} \hat{S}_{1\, p}^b \hat{S}_2^{pc} 
n_{\langle abc \rangle}. 
\end{equation} 
With all this we find that the spin terms in the black hole's
acceleration are given by 
\begin{equation} 
a_{1\, a}[\text{spin}] = \frac{3Gm_2}{c^2 b^3} \Bigl[ 
2 n_{\langle ac \rangle} \bigl( \hat{S}_{1\, b}^c + \hat{S}_{2\, b}^c \bigr) 
+ n_{\langle bc \rangle} \bigl( \hat{S}_1^{ac} + 2\hat{S}_2^{ac}
\bigr) \Bigr] u^b 
+ \frac{15 G m_2}{c^2 b^4} \hat{S}_{1\, p}^b \hat{S}_2^{pc} 
n_{\langle abc \rangle}. 
\label{acc_spin} 
\end{equation} 
The first group of terms, linear in the spins, is the well-known
expression for the leading-order spin-orbit acceleration of a
post-Newtonian body --- see Eq.~(9.245) of Poisson and Will, evaluated
with $\lambda = 1$ --- and the last term, bilinear in the spins, is
the leading-order spin-spin acceleration --- see Eq.~(9.190) of
Poisson and Will. As promised, we have verified that the acceleration
of Eq.~(\ref{acceleration}) reproduces the standard results of 
post-Newtonian theory. The spin acceleration $\bm{a}_2[\text{spin}]$
of the companion can be obtained directly from Eq.~(\ref{acc_spin}) by
switching the body labels and letting $\bm{n} \to -\bm{n}$, $\bm{u}
\to -\bm{u}$.    

To conclude this section we calculate the spin contributions to
$dA/dt$ and $dR^a/dt$, which are involved in the transformation from
the barycentric frame $(t,x^a)$ to the black-hole frame
$(\bar{t},\bar{x}^a)$. The calculation proceeds from Eqs.~(\ref{Adot})
and (\ref{Rdot}), it involves the same external potentials that were
computed previously, and it involves the Newtonian piece of $\E_{ab}$,
given by Eq.~(7.10) of Taylor and Poisson. After straightforward
manipulations we obtain 
\begin{equation} 
\frac{dA}{dt}[\text{spin}] = 0, \qquad 
\frac{dR_a}{dt}[\text{spin}] = -\frac{3Gm_2}{b^3} 
\Bigl( -\gam{d} \hat{S}_1^b + \hat{S}_2^b \Bigr) 
n_{\langle ab \rangle}. 
\label{Rdot_spin} 
\end{equation} 
These contributions are to be added to the nonspinning contributions
listed in Eq.~(7.12) of Taylor and Poisson. 

\section{Tidal moments for a circular binary} 
\label{sec:circular} 

In this section we continue our computation of the tidal moments
$\E_{ab}$ and $\B_{ab}$ when the black hole is a member of a two-body
system, and specialize the situation examined in the previous section
to circular motion.  We now assume that $b = \mbox{constant}$, and to
be consistent we take the spin vectors $\bm{S}_1$ and $\bm{S}_2$ to
be either aligned or anti-aligned with the orbital angular-momentum
vector. We describe the motion in terms of the orbital phase angle
$\phi := \omega t$, with $\omega$ denoting the orbital angular
velocity in the barycentric frame $(t,x^a)$ of the post-Newtonian
metric. We let the orbital motion take place in the $x$-$y$ plane of
the coordinate system, and in these coordinates we denote by 
$\bm{n} = [\cos\phi,\sin\phi,0]$ the unit vector that points
from the companion to the black hole. We also introduce the additional 
basis vectors $\bm{\phi} := [-\sin\phi,\cos\phi,0]$ and 
$\bm{\ell} := [0,0,1]$, with $\bm{\ell}$ pointing in the direction of
the orbital angular-momentum vector. In terms of the vectorial basis
we have $\bm{b} = b \bm{n}$, $\bm{u} = \omega b \bm{\phi}$,  
$\bm{S}_1 = S_1 \bm{\ell}$ and $\bm{S}_2 = S_2 \bm{\ell}$, with
$S_{1,2}$ positive (negative) when $\bm{S}_{1,2}$ is aligned
(anti-aligned) with $\bm{\ell}$. We resume the use of the
dimensionless spin variables $\chi_{1,2}$ defined by 
\begin{equation} 
S_1 = \frac{G m_1^2}{c} \chi_1, \qquad 
S_2 = \frac{G m_2^2}{c} \chi_2. 
\label{S_vs_chi} 
\end{equation} 
We shall work with the mass combinations $m := m_1 + m_2$ (total mass)
and $\eta := m_1 m_2/m^2$ (symmetric mass ratio).  

A concrete expression for the relative acceleration $\bm{a} :=
\bm{a}_1 - \bm{a}_2$ of the two-body system can be obtained from
Eqs.~(9.142) and (9.252) of Poisson and Will \cite{poisson-will:14}
(the spin-orbit terms must be evaluated with $\lambda = 1$). After
specialization to circular motion, which implies 
$\bm{a} = -b\omega^2 \bm{n}$, and substitution of
Eq.~(\ref{S_vs_chi}), the equations of motion produce the following
expression for the orbital velocity:   
\begin{equation} 
u^2 = (\omega b)^2 = \frac{Gm}{b} \biggl[ 1 
- (3-\eta) \frac{Gm}{c^2 b}
- \tilde{\chi} \biggl( \frac{Gm}{c^2 b} \biggr)^{3/2} 
+ O(c^{-4}) \biggr],  
\label{v_vs_r} 
\end{equation} 
where 
\begin{equation} 
\tilde{\chi} := \frac{m_1(2m_1+3m_2)}{m^2}\, \chi_1
+ \frac{m_2(2m_2+3m_1)}{m^2}\, \chi_2. 
\label{chi_tilde} 
\end{equation} 
This expression incorporates the Newtonian, post-Newtonian, and
spin-orbit terms in the relative acceleration, but it neglects the
spin-spin terms, which contribute at order $c^{-4}$ by virtue of the
scalings displayed in Eq.~(\ref{S_vs_chi}). The expression agrees with
Eq.~(4.5) of Ref.~\cite{kidder:95} 

It is a simple matter to specialize the expressions of
Eqs.~(\ref{Espin}) and (\ref{Bspin}) to circular motion. Making use of 
Eqs.~(\ref{spin-vector-tensor}) and (\ref{v_vs_r}), we obtain  
\begin{subequations} 
\begin{align} 
\E_{ab}[\text{spin}] &= \frac{6Gm_2}{b^3} (u/c)^3 
\frac{m_2}{m} \chi_2 \bigl( \phi_{\langle a b \rangle} 
+ 2 n_{\langle a b \rangle} \bigr) + O(c^{-5}), \\ 
\B_{ab}[\text{spin}] &= \frac{6Gm_2}{b^3} (u^2/c) 
\frac{m_2}{m} \chi_2\, \ell_{(a} n_{b)} + O(c^{-3}). 
\end{align} 
\end{subequations} 
Combining this with Eqs.~(7.19) and (7.20) of Taylor and Poisson
\cite{taylor-poisson:08}, we arrive at 
\begin{subequations} 
\label{EB_circ} 
\begin{align} 
\E_{ab} &= -\frac{3Gm_2}{b^3} \Biggl\{ \biggl[ 
1 - \frac{m_1+2m_2}{2m} (u/c)^2 
- \frac{4 m_2}{m} \chi_2 (u/c)^3 \biggr] n_{\langle ab \rangle} 
+ \biggl[ (u/c)^2 - \frac{2m_2}{m} \chi_2 (u/c)^3 \biggr] 
\phi_{\langle ab \rangle} \Biggr\} + O(c^{-4}), \\ 
\B_{ab} &= -\frac{6Gm_2}{b^3} u \biggl[ 
1 - \frac{m_2}{m} \chi_2 (u/c) \biggr] \ell_{(a} n_{b)} 
+ O(c^{-2}), 
\end{align} 
\end{subequations} 
complete expressions for the barycentric tidal moments, expanded 
through $1.5\PN$ order. 

The transformation of the tidal moments from the barycentric frame to
the black-hole frame involves the rotation tensor ${\cal N}_{ab}(t)$ of
Eq.~(\ref{N_def}), which is determined by $A(t)$ and $\bm{R}(t)$ as
obtained by integrating Eqs.~(\ref{Adot}) and (\ref{Rdot}). The spin
contributions to $dA/dt$ and $dR^a/dt$ were displayed in
Eq.~(\ref{Rdot_spin}), and again it is a simple matter to specialize
these results to circular motion. After inclusion of Eqs.~(7.21) and
(7.22) of Taylor and Poisson, we find that 
\begin{subequations} 
\begin{align} 
\frac{dA}{dt} &= \frac{(2m_1+3m_2) m_2}{2m^2} u^2 
+ O(c^{-2}),\\ 
\frac{dR^a}{dt} &= -\frac{Gm_2}{b^2} u \biggl[ 
\frac{4m_1+3m_2}{2m} + \biggl( \gam{d} \frac{m_1}{m} \chi_1 
- \frac{m_2}{m} \chi_2 \biggr) (u/c) \biggr]\ell^a + O(c^{-2}). 
\end{align}
\end{subequations} 
These equations imply that 
\begin{equation} 
t = \biggl[ 1 + \frac{(2m_1+3m_2) m_2}{2m^2} (u/c)^2  
+ O(c^{-4}) \biggr] \bar{t} 
\label{t_vs_tbar} 
\end{equation} 
and $c^{-2} \bm{R}(t) = -(\Omega t) \bm{\ell}$, where 
\begin{equation} 
\Omega := \sqrt{\frac{Gm}{b^3}} \biggl[ 
\frac{(4m_1+3m_2)m_2}{2m^2} (u/c)^2 
+ \frac{m_2}{m} \biggl( \gam{d} \frac{m_1}{m} \chi_1 
- \frac{m_2}{m} \chi_2 \biggr) (u/c)^3 + O(c^{-4}) \biggr]  
\label{Omega} 
\end{equation} 
is the precessional angular frequency of the black-hole frame 
$(\bar{t},\bar{x}^a)$ relative to the barycentric frame $(t,x^a)$. 

With these results it is easy to show that ${\cal N}_{ab}(t)$ has the
following effect on the basis vectors associated with the circular
motion: 
\begin{equation} 
{\cal N}_a^{\ c}\, n_c = \bar{n}_a := n_a(\phi-\Omega t), \qquad  
{\cal N}_a^{\ c}\, \phi_c = \bar{\phi}_a := \phi_a(\phi-\Omega t),
\qquad 
{\cal N}_a^{\ c}\, \ell_c = \bar{\ell}_a := \ell_a. 
\end{equation} 
We see that the dependence of each vector on the orbital phase 
$\phi = \omega t$ must be replaced by a dependence on 
$\phi - \Omega t = (\omega-\Omega) t$. If we also incorporate the 
transformation of the time coordinate, we find that
\begin{equation} 
\bar{\bm{n}} = [\cos\bar{\phi}, \sin\bar{\phi},0], \qquad 
\bar{\bm{\phi}} = [-\sin\bar{\phi}, \cos\bar{\phi},0], \qquad 
\bar{\bm{\ell}} = [0,0,1], 
\end{equation} 
where $\bar{\phi} = \bar{\omega} \bar{t}$ is now the phase of the
tidal field, with $\bar{\omega} := (\omega-\Omega)(t/\bar{t})$
describing the angular frequency of the tidal moments as measured in
the black-hole frame. Involving Eqs.~(\ref{v_vs_r}),
(\ref{t_vs_tbar}), and (\ref{Omega}), we see that a concrete
expression for the tidal angular frequency is  
\begin{equation} 
\bar{\omega} = \sqrt{\frac{Gm}{b^3}} \biggl[ 1 
- \frac{1}{2} (3+\eta) (u/c)^2 - \frac{1}{2} \bar{\chi} (u/c)^3 
+ O(c^{-4}) \biggr],  
\label{omegabar} 
\end{equation} 
where 
\begin{equation} 
\bar{\chi} := \frac{m_1}{m^2} \bigl[ 2 m_1 
+ (3 + 2\gam{d}) m_2 \bigr] \chi_1
+ 3 \eta \chi_2. 
\label{chibar} 
\end{equation} 
In view of the discussion of Sec.~\ref{sec:hor_lock}, which provides a 
precise definition of the black-hole frame in terms of the null
generators of the deformed event horizon, it is appropriate to
substitute $\gam{d} = -1$ in this expression.  

With all this we find that the tidal moments in the black-hole frame
take the same form as in Eq.~(\ref{EB_circ}), but with each vector
transformed in the way just described. Concretely, we have 
\begin{subequations} 
\label{EBbar_circ} 
\begin{align} 
\bar{\E}_{ab} &= -\frac{3Gm_2}{b^3} \Biggl\{ \biggl[ 
1 - \frac{m_1+2m_2}{2m} (u/c)^2 
- \frac{4 m_2}{m} \chi_2 (u/c)^3 \biggr] \bar{n}_{\langle ab \rangle}  
+ \biggl[ (u/c)^2 - \frac{2m_2}{m} \chi_2 (u/c)^3 \biggr] 
\bar{\phi}_{\langle ab \rangle} \Biggr\} + O(c^{-4}), \\ 
\bar{\B}_{ab} &= -\frac{6Gm_2}{b^3} u \biggl[ 
1 - \frac{m_2}{m} \chi_2 (u/c) \biggr] \bar{\ell}_{(a} \bar{n}_{b)} 
+ O(c^{-2}),  
\end{align} 
\end{subequations} 
with each basis vector $\bar{\bm{n}}$, $\bar{\bm{\phi}}$, and
$\bar{\bm{\ell}}$ expressed in terms of 
$\bar{\phi} = \bar{\omega} \bar{t}$.  

The individual components of $\bar{\E}_{ab}$ are best displayed in 
terms of the combinations $\bar{\E}_0 :=\frac{1}{2}
(\bar{\E}_{11} + \bar{\E}_{22})$, $\bar{\E}_{1c} := \bar{\E}_{13}$, 
$\bar{\E}_{1s} := \bar{\E}_{23}$, $\bar{\E}_{2c} := \frac{1}{2}
(\bar{\E}_{11} - \bar{\E}_{22})$, and $\bar{\E}_{2s} := \bar{\E}_{12}$. We
have that the nonvanishing components are 
\begin{subequations} 
\label{Ebar_components} 
\begin{align}   
\bar{\E}_0 &= -\frac{Gm_2}{2b^3} \biggl[ 1 
+ \frac{m_1}{2m} (u/c)^2 - 6 \frac{m_2}{m} \chi_2 (u/c)^3 
+ O(c^{-4}) \biggr], \\ 
\bar{\E}_{2c} &= -\frac{3Gm_2}{2b^3} \biggl[ 1 
- \frac{3m_1+4m_2}{2m} (u/c)^2 - 2 \frac{m_2}{m} \chi_2 (u/c)^3 
+ O(c^{-4}) \biggr] \cos 2\bar{\phi}, \\ 
\bar{\E}_{2s} &= -\frac{3Gm_2}{2b^3} \biggl[ 1 
- \frac{3m_1+4m_2}{2m} (u/c)^2 - 2 \frac{m_2}{m} \chi_2 (u/c)^3 
+ O(c^{-4}) \biggr] \sin 2\bar{\phi}. 
\end{align} 
\end{subequations} 
With similar definitions holding for $\bar{\B}_{ab}$, we have that its
nonvanishing components are 
\begin{subequations} 
\label{Bbar_components} 
\begin{align} 
\bar{\B}_{1c} &= -\frac{3Gm_2}{b^3} u \biggl[ 
1 - \frac{m_2}{m} \chi_2 (u/c) + O(c^{-2}) \biggr] \cos\bar{\phi}, \\ 
\bar{\B}_{1s} &= -\frac{3Gm_2}{r^3} u \biggl[ 
1 - \frac{m_2}{m} \chi_2 (u/c) + O(c^{-2}) \biggr] \sin\bar{\phi}. 
\end{align} 
\end{subequations} 
These results were already displayed in Sec.~\ref{sec:intro} --- refer
back to Eq.~(\ref{EB_components}) and the following equations. In the
expressions listed in Sec.~\ref{sec:intro}, the overbars were omitted
on the tidal moments, and the tidal phase was expressed more generally 
as $\bar{\phi} = \bar{\omega}(\bar{t}-\bar{t}_0)$ and then rewritten
as $\bar{\phi} = \bar{\omega}(v-v_0)$ in terms of the advanced-time
coordinate $v$. The expressions also incorporated the phase shift
$\delta \phi = \frac{8}{3} (m_1/m) (u/c)^3$ to be calculated in
Sec.~\ref{sec:Edot}.   

\section{Inclusion of time-derivative terms} 
\label{sec:Edot} 

A complete determination of the tidal moments at $1.5\PN$ order should   
incorporate terms in the metric that involve 
$\dot{\E}_{ab} := d\E_{ab}/dv$. These terms were neglected in
Eq.~(\ref{metric}), but they were present in the metrics constructed
by Poisson \cite{poisson:05} and Poisson and Vlasov
\cite{poisson-vlasov:10}. The relevant pieces of the metric
perturbation, which include both the $\E_{ab}$ and $\dot{\E}_{ab}$
terms, are given by      
\begin{subequations}  
\label{Edot_LC} 
\begin{align} 
p_{vv} &= -r^2 \eq_1\, \Eq 
+ \frac{1}{3} r^3 \eq_2\, \Edotq, \\  
p_{vA} &= -\frac{2}{3} r^3 \eq_4\, \Eq_A 
+ \frac{1}{3} r^4 \eq_5\, \Edotq_A, \\ 
p_{AB} &= -\frac{1}{3} r^4 \eq_7\, \Eq_{AB}
+ \frac{5}{18} r^5 \eq_8\, \Edotq_{AB},  
\end{align} 
\end{subequations} 
where $\Edotq$, $\Edotq_A$, and $\Edotq_{AB}$ are tidal potentials
constructed from $\dot{\E}_{ab}$; the radial functions $\eq_1$,
$\eq_4$, and $\eq_7$ were listed in Table~\ref{tab:radial}, while
$\eq_2$, $\eq_5$, and $\eq_8$ are displayed in
Table~\ref{tab:Edot}. It should be noted that the residual freedom of
the light-cone gauge and the freedom to redefine the tidal moments
according to $\E_{ab} \to \E_{ab} + p M \dot{\E}_{ab}$, where $p$ is
an arbitrary number, were exploited to ensure that $\eq_2$, $\eq_5$,
and $\eq_8$ all go to zero when $r=2M$; this implies that the horizon
metric of Eq.~(\ref{hor_metric}) does not acquire a contribution from
$\dot{\E}_{ab}$. The $\dot{\E}_{ab}$ terms of Eq.~(\ref{Edot_LC}) can
be added to the metric of Eq.~(\ref{metric}) to obtain a more complete
description of the geometry of a tidally deformed, slowly rotating
black hole.     

That the $\dot{\E}_{ab}$ terms should be incorporated in a
$1.5\PN$ calculation of the tidal moments can be seen from the
following argument. Equation~(\ref{Edot_LC}) reveals that the leading
tidal term in $g_{vv}$ is proportional to $r^2 \E_{ab}$, while the
time-derivative term is proportional to $r^3 \dot{\E}_{ab}$. If we take
$r$ to be of the same order of magnitude as $M$ and denote by $\tau$
the time scale associated with changes in the tidal environment, we
find that the time-derivative term is of order $M/\tau$ relative to
the leading term. With $\tau$ of order $\sqrt{b^3/M}$, where $b$ is
the distance scale to the external matter, we find that
$M/\tau \sim (M/b)^{3/2} \sim u^3$, where $u \sim b/\tau$ is the
velocity scale. Restoring factors of $c$, we have found that the
time-derivative term is smaller than the leading term by a factor of
order $(u/c)^3$, and that it therefore represents a $1.5\PN$
correction to the metric.  

There were, nevertheless, good reasons to neglect the $\dot{\E}_{ab}$
terms until now. These were not included in the metric of
Eq.~(\ref{metric}) because unless $\chi$ is extremely small, they are
negligible compared to the $\chi\E_{ab}$ terms when $r$ is comparable
to $M$, which was the prevailing context until the very end of
Sec.~\ref{sec:harmonic}. But because they create $1.5\PN$ corrections
to the tidal moments, we now incorporate the $\dot{\E}_{ab}$ in a more
complete description of the metric, and an improved determination of
the tidal moments.     

%\begingroup
%\squeezetable
\begin{table}
\caption{Radial functions associated with the $\dot{\E}_{ab}$ terms in
  the metric; $f := 1-2M/r$.}   
\begin{ruledtabular} 
\begin{tabular}{l} 
$\eq_2 = f \bigl[ 1 + \frac{M}{2r}(5 + 12 \ln \frac{r}{2M}) 
- \frac{M^2}{r^2} (27 + 12 \ln \frac{r}{2M}) 
+ \frac{14M^3}{r^3} + \frac{12M^4}{r^4} \bigr]$ \\ 
$\eq_5 = f \bigl[ 1 + \frac{M}{3r}(13 + 12 \ln \frac{r}{2M}) 
- \frac{10M^2}{r^2} - \frac{12M^3}{r^3} 
- \frac{8M^4}{r^4} \bigr]$ \\ 
$\eq_8 = 1 + \frac{4M}{5r}(4 + 3 \ln \frac{r}{2M}) 
- \frac{36M^2}{5r^2} 
- \frac{8M^3}{5r^3} (7 + 3\ln \frac{r}{2M}) 
+ \frac{48M^4}{5r^4}$ \\
{ } \\ 
$z_v = \frac{2(r-2M)M(35r^3-12Mr^2+24M^2r-8M^3)}{105 r^5} 
\ln\frac{r}{2M} 
+ \frac{(r-2M)(150r^4+415Mr^3+342M^2r^2-3540M^3r-288M^4)} 
{630 r^5}$ \\ 
$z_r = -\frac{2M(35r^3-12Mr^2+24M^2r-8M^3)}{105r^4} 
\ln\frac{r}{2M} 
- \frac{405r^5+34250Mr^4-103036M^2r^3+63232M^3r^2+6584M^4r
+1152M^5}{1260 r^4 (r-2M)}$ \\ 
$z = \frac{2M(r-2M)^2}{3r^2(r-M)} \ln\frac{r}{2M}  
+ \frac{5r^5+317Mr^4-1036M^2r^3+1056M^3r^2-272M^4r-48M^5} 
{18 r^4(r-M)}$ \\ 
{ } \\ 
$w_{vv} = \frac{2M(r-2M)^2}{r^3} \ln\frac{r}{2M} 
+ \frac{9r^5-12Mr^4-582M^2r^3+1662M^3r^2-1072M^4r+28M^5}
{9r^5}$ \\ 
$w_{vr} = -\frac{2(r-2M)M(35r^2-8Mr+8M^2)}{35r^4} \ln\frac{r}{2M} 
- \frac{405r^6-630Mr^5-18870M^2r^4+51654M^3r^3-28304M^4r^2 
-2812M^5r-192M^6}{315 r^5 (r-2M)}$ \\ 
$w_{rr} = \frac{M(140r^2-32Mr+32M^2)}{35r^3} \ln\frac{r}{2M}
+ \frac{2(405r^6+24885Mr^5-120930M^2r^4+171634M^3r^3 
-64144M^4r^2-2812M^5r-192M^6)}{315 r^4 (r-2M)^2}$ \\ 
$w_v = \frac{16M^2(r-2M)(3r^2-6Mr+2M^2)}{105 r^5} \ln\frac{r}{2M} 
+ \frac{(r-2M)(60r^5-125Mr^4-1432M^2r^3+3672M^3r^2-2832M^4r 
+ 552M^5)}{315 r^5 (r-M)}$ \\ 
$w_r = \frac{2M^2(81r^4-114Mr^3+4M^2r^2+80M^3r-16M^4)}
{105 r^4 (r-M)^2} \ln\frac{r}{2M} 
- \frac{60r^7-6290Mr^6+27833M^2r^5-42192M^3r^4+21018M^4r^3 
+5192M^5r^2-6340M^6r+1104M^7}{315 r^4 (r-M)^2 (r-2M)}$ \\ 
$\bar{w} = \frac{2M(r-2M)^2}{r^2(r-M)} \ln\frac{r}{2M}
+ \frac{9r^5+960Mr^4-3498M^2r^3+4066M^3r^2-1532M^4r+28M^5} 
{9r^4(r-M)}$ \\  
$w = \frac{2M^2(3r^2-6Mr+2M^2)}{r^3(r-M)} \ln\frac{r}{2M} 
- \frac{M(153r^3-492Mr^2+538M^2r-188M^3)}{9r^3(r-M)}$ 
\end{tabular}
\end{ruledtabular} 
\label{tab:Edot} 
\end{table} 
%\endgroup

The metric perturbation of Eq.~(\ref{Edot_LC}) can be recast in the
harmonic gauge by following the methods described in
Sec.~\ref{sec:harmonic}. The calculations are simpler here, because
the gauge transformation is to be carried out to first order only in
the perturbation. A complication arises, however, because the time
dependence of the tidal moments $\E_{ab}(v)$ must now be taken into
account. Recalling Eq.~(\ref{gauge_E}), the generating vector field is
expressed as   
\begin{equation} 
\xi_v = -\frac{1}{3} r^3 f \Eq + r^4 z_v\, \Edotq, \qquad 
\xi_r = \frac{1}{3} r^3 \Eq + r^4 z_r\, \Edotq, \qquad 
\xi_A = -\frac{r^5 f^2}{3(r-M)} \Eq_A + r^4 z\, \Edotq_A, 
\end{equation} 
where $z_v$, $z_r$, and $z$ are radial functions determined by
integrating the differential equations issued from the harmonic 
conditions of Eq.~(\ref{harm1}). The solutions are listed in
Table~\ref{tab:Edot}, for some suitable choice of integration
constants. After inserting the gauge vector in
Eq.~(\ref{gauge_transf}) and truncating to first order, we find that
the harmonic-gauge form of the perturbation is given by 
\begin{subequations} 
\begin{align} 
p_{vv} &= -r^2 \eq_{vv}\, \Eq 
+ r^3 w_{vv}\, \Edotq, \\ 
p_{vr} &= r^2 \eq_{vr}\, \Eq 
+ r^3 w_{vr}\, \Edotq, \\
p_{rr} &= -2r^2 \eq_{rr}\, \Eq  
+ r^3 w_{rr}\, \Edotq, \\
p_{vA} &= r^4 w_v\, \Edotq, \\ 
p_{rA} &= -r^3 \eq_{r}\, \Eq_A 
+ r^3 w_r\, \Edotq_A, \\ 
p_{AB} &= - r^4 \ebarq\, \Omega_{AB} \Eq - r^4 \eq\, \Eq_{AB}   
+ r^5 \bar{w}\, \Omega_{AB} \Edotq + r^5 w\, \Edotq_{AB}, 
\end{align} 
\end{subequations} 
where the various $\eq$-functions are listed in
Table~\ref{tab:radial_harm}, while the $w$-functions are listed in
Table~\ref{tab:Edot}. 

We next follow the procedure described near the end of
Sec.~\ref{sec:harmonic}: We carry out a transformation to Cartesian 
harmonic coordinates, and express the metric perturbation as a
post-Newtonian expansion truncated to $1.5\PN$ order. There is no need
to go through the details again here, but one aspect to be careful about is
that the transformation to harmonic coordinates involves expressing
$\E_{ab}$ as a function of $v = t + r \Delta$, where $\Delta := 1 +
(2M/r) \ln(r/2M - 1)$. Because the time dependence of the tidal
moments is slow, we can write this as  
\begin{equation} 
\E_{ab}(t+r\Delta) = \E_{ab}(t) + r \Delta \dot{\E}_{ab}(t) + \cdots 
\end{equation}  
and express $p_{vv}$ (say) as 
\begin{equation} 
p_{vv} = -r^2 \eq_{vv}\, \Eq(t)  
+ r^3 \bigl( w_{vv} - \Delta \eq_{vv} \bigr) \, \Edotq(t). 
\end{equation} 
Completing the calculations, and inserting the rotation-tidal
couplings that were considered previously, we find that the metric of
Eq.~(\ref{metric_harm_PN2}) becomes   
\begin{subequations} 
\label{metric_harm_Edot} 
\begin{align} 
g_{tt} &= -1 + \frac{2GM}{c^2 \bar{r}}
-2 \biggl(\frac{GM}{c^2 \bar{r}} \biggr)^2 
- \frac{1}{c^2} \biggl( 1 - \frac{2GM}{c^2 \bar{r}} \biggr) 
\E_{ab} x^a x^b 
+ \frac{8}{3} \frac{GM}{c^5} \dot{\E}_{ab} x^a x^b  
+ \frac{2}{c^4} \B_{ap} \hat{S}^p x^a , \\ 
g_{ta} &= \frac{2G}{c^3} \frac{(\bm{x} \times \bm{S})_a}{\bar{r}^3} 
+ \frac{2}{3c^3} \epsilon_{abp} \B^p_{\ c} x^b x^c  
- \frac{\gam{d}}{c^3} \epsilon_{abp} \E^p_{\ q} \hat{S}^q x^b 
\end{align} 
\end{subequations} 
when time-derivative terms are properly incorporated. It should 
be noted that the tidal moments are now expressed as functions of
harmonic time $t$, and that $g_{ta}$ is actually unchanged relative to
Eq.~(\ref{metric_harm_PN2}).  

The metric of Eq.~(\ref{metric_harm_Edot}) can be rewritten in a form
identical to Eq.~(\ref{metric_harm_PN2}) by simply introducing the new
tidal moments 
\begin{equation} 
\E^\sharp_{ab} := \E_{ab} - \frac{8}{3} \frac{GM}{c^3} \dot{\E}_{ab}. 
\label{Esharp1} 
\end{equation} 
With this redefinition, the calculations presented in
Secs.~\ref{sec:tidal}, \ref{sec:2body}, \ref{sec:circular} proceed
completely unchanged. The tidal moments determined in these sections
are then $\E^\sharp_{ab}$, and the true moments $\E_{ab}$ are
recovered by inverting Eq.~(\ref{Esharp1}), 
\begin{equation} 
\E_{ab} = \E^\sharp_{ab} + \frac{8}{3} \frac{GM}{c^3}
\dot{\E}^\sharp_{ab} + O(c^{-6}).  
\label{Esharp2} 
\end{equation} 
In this way Eq.~(\ref{EpN}) becomes 
\begin{equation} 
\E_{ab} = \E^\sharp_{ab} - \frac{8}{3} \frac{GM}{c^3} 
\Bigl( \partial_{tab} U_{\rm ext} + u^c \partial_{abc} U_{\rm ext}
\Bigr) + O(c^{-4}), 
\label{Esharp3} 
\end{equation} 
where $\E^\sharp_{ab}$ is the expression previously displayed in 
Eq.~(\ref{EpN}). Specializing to a two-body system, and implementing
the change of notation described at the beginning of
Sec.~\ref{sec:2body}, this is   
\begin{equation} 
\E_{ab} = \E^\sharp_{ab} + 40 \frac{G^2 m_1 m_2}{c^3} 
\frac{ n_{\langle a b c \rangle} u^c }{b^4} + O(c^{-4}), 
\label{Esharp4} 
\end{equation} 
where $\E^\sharp_{ab}$ is now the expression shown in
Eq.~(7.10) of Taylor and Poisson \cite{taylor-poisson:08}, added to
the spin terms displayed in Eq.~(\ref{Espin}). Specializing further to
a two-body system in circular motion, the true tidal moments are given 
by  
\begin{equation} 
\E_{ab} = \E^\sharp_{ab}  - 16 \frac{Gm_2}{b^3} 
\frac{m_1}{m} (u/c)^3 n_{(a} \phi_{b)}, 
\label{Esharp5} 
\end{equation} 
where $\E^\sharp_{ab}$ now stands for the expression of
Eq.~(\ref{EB_circ}).  
 
For a binary system in circular motion, Eq.~(\ref{Esharp2}) implies
that $\E_{ab}$ and $\E^\sharp_{ab}$ differ only by a constant shift in
the phase. The equation can indeed be re-written as 
\begin{equation} 
\E_{ab} = \E^\sharp_{ab} + \frac{8}{3} \frac{GM\omega}{c^3}
\partial_\phi \E^\sharp_{ab} + O(c^{-6}), 
\label{Esharp6} 
\end{equation} 
and this can in turn be expressed as 
\begin{equation} 
\E_{ab}(\phi) = \E^\sharp_{ab}(\phi + \delta\phi), \qquad 
\delta\phi := \frac{8}{3} \frac{m_1}{m} (u/c)^3 + O(c^{-5}). 
\label{Esharp7} 
\end{equation} 
It is easy to show that this equation is compatible with
Eq.~(\ref{Esharp5}). This constant phase shift is largely
uninteresting, but the more general situation described by
Eq.~(\ref{Esharp4}) can indeed describe interesting, time-dependent
phasing effects.  

\begin{acknowledgments} 
I am grateful to the Institut d'Astrophysique de Paris, where part of
this work was completed, for its warm hospitality.  This work was
supported by the Natural Sciences and Engineering Research Council of
Canada.     
\end{acknowledgments}    

\bibliography{../bib/master} 
\end{document}